\documentclass{article}

\usepackage{PRIMEarxiv}

\usepackage[utf8]{inputenc} 
\usepackage[T1]{fontenc}    

\usepackage{url}            
\usepackage{booktabs}       
\usepackage{nicefrac}       
\usepackage{microtype}      
\usepackage{lipsum}
\usepackage{fancyhdr}       
\usepackage{graphicx}       
\graphicspath{{media/}}     

\usepackage{graphicx}%
\usepackage{multirow}%
\usepackage{amsmath,amssymb,amsfonts}%
\usepackage{amsthm}%
\usepackage{mathrsfs}%
\usepackage[title]{appendix}%
\usepackage{xcolor}%
\usepackage{textcomp}%
\usepackage{manyfoot}%
\usepackage{booktabs}%
\usepackage{algorithm}%
\usepackage{algorithmicx}%
\usepackage{algpseudocode}%
\usepackage{listings}%
\usepackage[numbers,sort&compress]{natbib}  
\usepackage[colorlinks=true,
            linkcolor=blue,
            citecolor=blue,
            urlcolor=blue]{hyperref}
            
\usepackage{dsfont}

\usepackage{adjustbox}

\theoremstyle{thmstyleone}%
\newtheorem{theorem}{Theorem}
\theoremstyle{thmstyletwo}%

\theoremstyle{thmstylethree}%
\newtheorem{definition}{Definition}%

\title{Quantile Forecast Matching with a Bayesian Quantile Gaussian Process Model
\thanks{\textit{\underline{Citation}}:
\textbf{DOI:2502.06605.}
}
}

\author{
  Spencer Wadsworth \\
  University of Connecticut \\
  Storrs, CT 06269\\
  \texttt{spencer.wadsworth@uconn.edu} \\
   \And
  Jarad Niemi \\
  Iowa State University \\
  Ames, IA 50011\\
  \texttt{niemi@uconn.edu} \\
}

\begin{document}
\maketitle

\begin{abstract}
A set of probabilities along with corresponding quantiles are often used to 
define predictive distributions or probabilistic forecasts. These quantile 
predictions offer easily interpreted uncertainty of an event, 
and quantiles are generally straightforward to estimate using standard 
statistical and machine learning methods. However, compared to a distribution 
defined by a probability density or cumulative distribution function, a set of 
quantiles has less distributional information. When given estimated 
quantiles, it may be desirable to estimate a fully defined continuous 
distribution function. Many researchers do so to make evaluation or 
ensemble modeling simpler. Most existing methods for fitting a distribution to 
quantiles lack accurate representation of the inherent uncertainty from 
quantile estimation or are limited in their applications. In this manuscript, we 
present a Gaussian process model, the quantile Gaussian process, which is based 
on established asymptotic results
of quantile functions and sample quantiles, to construct 
a probability distribution given estimated quantiles. A Bayesian application of 
the quantile Gaussian process is evaluated for parameter inference and 
distribution approximation in simulation studies. The quantile Gaussian 
process is used to approximate the distributions of quantile forecasts from the 
2023-24 US Centers for Disease Control collaborative flu forecasting 
initiative. The simulation studies and data analysis show that the quantile 
Gaussian process leads to accurate inference on model parameters, estimation of 
a continuous distribution, and uncertainty quantification of sample quantiles.
The method being Bayesian also allows for the use of prior information when 
approximating a distribution given quantiles.
\end{abstract}


\section{Introduction} \label{seq:intro}

The use of quantiles in statistical modeling and in reporting inferential or 
predictive uncertainty is widespread, and reporting several quantiles or 
predictive intervals is common for probabilistic forecasting 
\cite[]{gneiting2023model}. Quantiles can be easier to interpret than 
statistical model parameters and are often used to define confidence or 
prediction intervals. Thus with multiple quantiles for a particular outcome, 
one has a measure of uncertainty.
Estimating quantiles in the presence of covariates via 
quantile regression is a common
statistical and machine learning strategy, and where parametric models are 
complicated or nonexistent quantiles may be easier to estimate via machine 
learning methods \cite[]{martin2022direct,chung2021beyond, koenker2017quantile, 
koenker1978regression}. In estimating multiple quantiles, quantile regression 
often provides a tradeoff with parametric modeling in that for some problems 
quantile regression may be easier to perform but that predicted quantiles lack 
the detailed information of a fully defined predictive distribution 
\cite[]{pohle2020murphy}. 
Perhaps for data privacy reasons, data quantiles are often reported as 
summaries of the data.
Census or medical data, which can be very large or personal to the subjects, 
may be published as summary or aggregate data including percentiles (quantiles) 
and medians 
\cite[]{simpson2023interpolating,cdc2022growthcharts,nirwan2020bayesian}. In 
collaborative forecast initiatives and competitions, probabilistic forecasts 
are often submitted as a set of predictive quantiles 
\cite[]{gneiting2023model,hong2016probabilistic}. In recent years, disease 
outbreak forecast hubs require that all forecasts submitted by outside 
participants be represented by  several predictive intervals or quantiles. 
This standardized representation allows for straightforward forecast scoring 
and ensemble building 
\cite[]{mathis2024evaluation,mathis2023flusight,Cramer2022-hub-dataset,
cramer2022evaluation,sherratt2023predictive,bracher2021evaluating}.
 
A set of quantiles may provide distributional information for an event, but it 
is not as informative as a distribution defined by a cumulative distribution 
function (CDF), the inverse-CDF or quantile function QF, or a probability 
density/mass function (PDF), and quantiles provide no distribution tail 
information --information for values below the smallest quantile or above the 
largest quantile. Another drawback for using quantiles to define a distribution 
is that many tools for evaluating or scoring continuous distributions require 
a CDF or PDF
\cite[]{gneiting2007strictly,gneiting2014probabilistic}. Likewise combining 
distributions into an ensemble distribution is commonly 
done by aggregating multiple QFs or 
CDFs/PDFs, the latter being possible only when a CDF/PDF is available 
\cite[]{gneiting2005calibrated,wang2023forecast}.

\begin{figure}[hbt!]
    \centering
    \includegraphics[scale=.5]{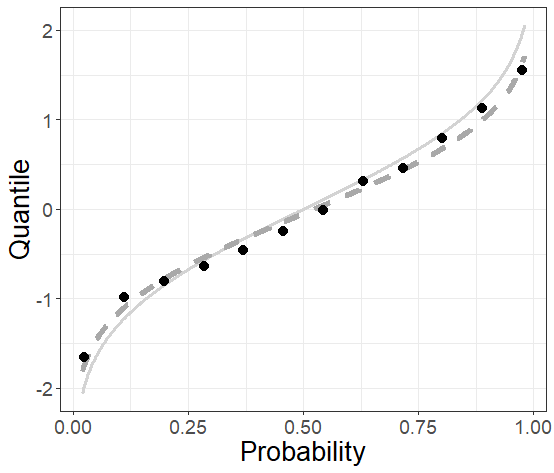}
    \caption{Points representing 12 quantiles ($y$-axis) for given probability 
    values ($x$-axis) estimated from a random sample of size 100 from a 
    standard normal distribution with the true quantile function (QF) 
    (solid grey). The estimated QF (dashed grey) was fit by selecting the mean 
    and standard deviation 
    parameters which minimize the least squares distance between the quantiles 
    and the QF.}
    \label{fig:quant_match_example}
\end{figure}

Fitting quantiles to recover or estimate a continuous distribution is done in 
many fields, often for the purpose of evaluating forecasts using rules that 
require a CDF or PDF \cite[]{simpson2023interpolating, gerding2023evaluating}. 
Fitting may also be done to allow combining multiple forecasts using aggregation 
methods that require CDFs or PDFs 
\cite[]{gyamerah2020probabilistic,li2019combining,baran2018combining,
bogner2017combining,he2016short,gneiting2005calibrated}. 
An example of fitting quantiles is given in figure 
\ref{fig:quant_match_example}. The 12 points in the figure are quantiles 
estimated from a random sample of size 100 from a standard normal distribution 
for given probabilities. The quantile function of the standard normal 
distribution is represented by the solid grey line in the plot. 
The dashed grey line is the fitted QF of a normal 
distribution that was estimated by selecting the mean and standard deviation 
parameters which minimize the least squares distance between the estimated 
quantiles and the QF.

Hereafter we refer to estimating a continuous distribution by fitting 
quantiles as quantile matching (QM).
\cite{sgouropoulos2015matching} performed QM by minimizing the mean square 
difference between quantiles of a response variable and a linear combination 
of quantiles of covariates. Selecting distribution parameters which minimize 
the mean square error
between quantiles and a QF is a common QM method 
\cite[]{dilger2022distributions,li2019combining, belgorodski2017quantilemse}, 
kernel density estimation and spline interpolation are common nonparametric 
QM methods 
\cite[]{gerding2023evaluating,gyamerah2020probabilistic,he2016short}, 
and \cite{keelin2016metalog} introduced a semiparametric method based on 
defining a flexible QF to fit quantiles. An issue shared by these methods is 
that they do not adequately account for the uncertainty inherent from the 
estimation of quantiles.  \cite{nirwan2020bayesian} introduced a QM model 
based on the definition of sample quantiles as order statistics and formulated
a model likelihood for the quantiles as the joint PDF of multiple order 
statistics. This model 
provides exact inference for quantile uncertainty, but relies on the CDF and 
PDF of the distribution, making it less practical for fitting quantile defined 
distributions which have a QF but for which the CDF and PDF either do not exist 
in closed form or are difficult to evaluate 
\cite[]{perepolkin2023tenets,joiner1971some,tukey1960practical}.

In this manuscript we introduce a novel model, the quantile Gaussian process 
(QGP) which is used specifically for QM.
The context for using the QGP is that the 
available data is a set of quantiles estimated for a given set of 
probabilities.
The QGP relies on established asymptotic 
theory underlying the relationship between the 
QF of a continuous distribution and sample quantiles 
\cite[]{parzen2004quantile, gilchrist2000statistical,hyndman1996sample,
walker1968note,cramer1951mathematical}. 
The QGP can provide accurate asymptotic inference of the quantile uncertainty, 
and it works well for QM when modeling quantile defined distributions. The 
QGP is fit using Bayesian methods and thus allows for using prior information
to influence the QM.
In section \ref{sec:quant_func} the QF of a continuous random variable and 
sample quantiles are each defined, and important properties and asymptotic 
theory are reviewed. In section \ref{sec:qgp_model} the QGP model for QM is 
introduced. Section \ref{sec:simulation_analyses} contains simulation studies 
illustrating the QGP's ability to make parameter inference and match the 
distribution from which the quantiles where estimated. Section 
\ref{sec:qm_analysis} contains an application of QM using the QGP model for 
quantile forecasts of the 2023-24 United States Centers for Disease Control and
Prevention
(CDC) flu forecasting competition, also known as FluSight. 
The analyses in sections \ref{sec:simulation_analyses} and
\ref{sec:qm_analysis} show that the QGP accurately estimates distribuions from
which the quantiles are estimated and while doing so accurately captures the 
uncertainty inherent in estimating quantiles. The QGP outperforms all
but one other competing methods in estimating the unknown distributions and
capturing uncertainty, and also allows for QM in situations where the other 
methods cannot be used.
Section 
\ref{sec:qm_conclusion} concludes this manuscript with a short summary and 
ideas for further development in QM modeling.

\section{Quantile function and sample quantiles} \label{sec:quant_func}

Along with the CDF and PDF, the QF is a defining function of a random variable. 
The QF, however, often receives less attention in standard statistical 
training than do the CDF and PDF \cite[]{parzen2004quantile}, and the CDF and 
PDF are used more often for modeling and inference.
This section contains the definition of the QF, important properties, the 
definition of sample quantiles, and key central limit theorems (CLTs) of sample 
quantiles on which the QGP is based.

\subsection{Quantile function definition and properties}

The QF is defined in definition \ref{def:qf}.

\begin{definition}
    \label{def:qf}
    For a cumulative distribution function $F: \mathbb{R} \rightarrow 
    \left[ 0,1 \right]$, 
    the quantile function $Q: \left[0,1\right] 
    \rightarrow \mathbb{R}$ is defined as 
    
    \begin{equation*}
    \label{eq:qf}
        Q\left(p\right) = F^{-1}\left(p\right) := 
        \emph{inf}\{x \in \mathbb{R} : F\left(x\right) \geq p\}, 
        \quad p \in \left[0,1\right]
    \end{equation*}
    An alternative definition not in terms of the cumulative distribution 
    function is a function 
    $Q: \left[0,1\right] \rightarrow \mathbb{R} \cup \{-\infty, \infty\}$, which is 
    nondecreasing and left-continuous. 
\end{definition}

For most statistical applications, it is useful to define the QF as the inverse 
of the CDF. However, it is sometimes the case that defining a distribution by 
a QF rather than a CDF is advantageous. For instance when a distribution is 
defined by a 
QF but has no closed form CDF or when simple QFs may be aggregated to define 
more complex distributions 
\cite[]{perepolkin2023tenets, gasthaus2019probabilistic, alvarez2023quantile}. 
This section continues with important properties of the QF and functions 
including the location-scale property, the quantile density function, the 
probability integral transform, and examples of quantile defined distributions.

\subsubsection{Location-scale property}
For a random variable with PDF $f_0\left(x\right)$ and 
QF $Q_0\left(p\right)$ for $p \in \left[0,1\right]$, a 
location-scale family takes the PDF form 
$\left(1/\sigma\right)f_0\left(\left(x - \mu\right)/\sigma\right)$ 
with location parameter 
$\mu \in \left( -\infty, \infty \right)$ and scale parameter 
$\sigma \in \left(0, \infty \right)$ \cite[]{casella2002statistical}. 
The QF of the 
location-scale family then takes the form $\mu + \sigma Q_0\left(p \right)$ 
\cite[]{parzen2004quantile}. By the addition and multiplication rules from 
section 3.2 of \cite{gilchrist2000statistical}, 
this is a valid QF. \cite{gilchrist2000statistical} outlines the 
standardization rule which states that if the random variable $Y$ with QF 
$Q_Y\left(p \right)$ has a standard distribution, that is the 
random variable $Y$ is 
centered (by a mean or median) at 0 and has a scale of 1, then a random 
variable with quantile function 
$Q_{T\left(Y \right)}\left(p \right) = \mu + \sigma Q_Y\left(p \right)$ is 
centered at $\mu$ and has scale $\sigma$.

The linear form for a QF is already inherent in some distributions families 
such as the normal and logistic distribution families, with respective QFs 
$\mu + \sigma \Phi^{-1} \left(p \right)$ and 
$\mu + \sigma \text{log}\left(p/ \left(1 - p \right) \right)$. For the
normal distribution, 
$\Phi^{-1}\left(p \right)$ is the QF of a standard normal distribution 
and $\mu$ and $\sigma$ are the mean and standard deviation of the 
location-scale transformed normal distribution. For the logistic distribution 
$\text{log}\left(p/\left(1-p \right) \right)$ 
is the QF of a logistic distribution with mean 0 and 
scale 1, and $\mu$ and $\sigma$ become the mean and scale parameters of the 
location-scale transformed logistic distribution. The linear form is convenient 
for quantile regression modeling and is key to the development of the metalog 
distribution family which was designed specifically for modeling quantiles 
\cite[]{keelin2016metalog}. 

\subsubsection{Quantile density function}

Besides the CDF, PDF, and QF, a  fourth infrequently mentioned defining function 
of a continuous random variable is the quantile density function (QDF). As the 
PDF is the derivative of the CDF, the QDF $q:[0,1] \rightarrow \mathbb{R}$ is 
the derivative of the QF defined as
\[
    q(p) = \frac{dQ(p)}{dp}.
\]
If $Q(p)$ is the inverse of a CDF $F(x)$, a calculus result shows that the 
reciprocal QDF $[q(p)]^{-1} = f(Q(p))$, or the PDF evaluated at the QF 
evaluated at $p$ 
\cite[]{perepolkin2023tenets, gilchrist2000statistical}. The QDF is important 
for establishing the CLTs in section \ref{sec:samp_quants} and for the QGP 
models introduced in section \ref{sec:qgp_model}.

\subsubsection{Probability integral transform}
An important application of the QF is using it to sample from continuous 
probability distributions via the probability integral transform (PIT). The 
PIT states that for any continuous random variable $Y$ with CDF $F_Y(y)$, the 
transformed random variable $X = F_Y(Y)$ is uniformly distributed on $(0,1)$, 
or $X \sim \text{Unif}(0,1)$. Thus if one can sample from a standard uniform 
distribution, one may also sample from a continuous distribution provided the 
QF of that distribution can be evaluated or reasonably approximated 
\cite[]{wilkinson2018stochastic}. The PIT is useful for assessing model fit via 
generalized residuals \cite[]{yang2024double, cox1968general} and the 
calibrating of probabilistic forecasts \cite[]{gneiting2007probabilistic}. 
Herein the PIT is important for developing one version of the QGP model, and it 
is used as part of a distance measure for assessing QM.

\subsubsection{Examples of quantile defined distributions} \label{sec:quant_def_dist}

An example of a quantile defined distribution family that lacks a CDF in closed 
form is the generalized lambda distribution (GLD) family 
\cite[]{ramberg1974approximate, perepolkin2023tenets}. A special case of the 
GLD family is the Tukey lambda distribution (TLD) family, a one parameter 
distribution family introduced by \cite{tukey1960practical}, 
\cite[]{joiner1971some}. The QF for the TLD is in (\ref{eq:tuk_lam}) and the 
QDF in (\ref{eq:tuk_lam_qd}).   

\begin{equation}
    \label{eq:tuk_lam}
    Q_{\lambda}(p) = \begin{cases} 
      \frac{1}{\lambda} \left[p^{\lambda} - (1 - p)^{\lambda} \right], & \lambda \neq 0 \\
      \text{log} \left(\frac{p}{1 - p} \right), & \lambda = 0
   \end{cases}
\end{equation}

\begin{equation}
    \label{eq:tuk_lam_qd}
    q_{\lambda}(p) = p^{\lambda - 1} + (1-p)^{\lambda -1} 
\end{equation}

Another distribution family defined by its quantile function but without a 
closed form CDF is the metalog distribution (MLD) family, a generalization of 
the logistic distribution family \cite[]{keelin2016metalog}. One version of the 
MLD is defined in (\ref{eq:mldq}).

\begin{equation}
    \label{eq:mldq}
    Q_{\textbf{a}}(p) = a_1 + a_3(p - 0.5) + a_2\text{log}\left(\frac{p}{1-p} \right)
\end{equation}
These two distributions are used in section \ref{sec:simulation_analyses} 
for comparing different quantile matching methods.

\subsection{Sample quantiles} \label{sec:samp_quants}

Using samples of a random variable to estimate quantiles is prevalent in 
statistical modeling. Given a random sample from some distribution, the sample 
quantile function may be defined in terms of the order statistics. 
\cite{hyndman1996sample} review several definitions of sample quantiles used in 
statistical packages and state definition \ref{def:sqf} as a general definition 
of the functions reviewed. 

\begin{definition}
For $n$ independent observations $Y_1, ..., Y_n$ from a distribution with 
corresponding order statistics $Y_{(1)}, ..., Y_{(n)}$, the sample quantile 
function $\hat{Q}_n(p)$ may be defined as
    \label{def:sqf}
    \begin{equation}
        \label{eq:quant_df}
        \hat{Q}_n(p) = (1- \gamma)Y_{(j)} + \gamma Y_{(j + 1)}
    \end{equation}
    where 
    \[\frac{j - m}{n} \leq p < \frac{j - m + 1}{n}
    \]
    for some $m \in \mathbb{R}$ and $0 \leq \gamma \leq 1$ where 
    $\gamma$ is some function of $j = \lfloor pn + m \rfloor$,  
    $g = pn + m - j$, and $\lfloor \cdot \rfloor$ is the floor function.
\end{definition}

Sample quantiles are used to give a probability summary of a dataset. Sample 
quantiles of Markov chain Monte Carlo draws from a Bayesian posterior 
distribution are used to analyze posterior distributions, for example by using 
quantiles to form credible intervals. 
Well known asymptotic results of sample quantiles are presented below, and it 
may be noted that asymptotic results for quantile regression quantiles are 
similar \cite[]{kocherginsky2005practical, koenker1978regression}. 

\subsubsection{Sample quantile central limit theorem} 

The asymptotic distribution of a set of sample quantiles is given in 
(\ref{eq:qclt}), and it is the basis for the QGP model for QM in section 
\ref{sec:qgp_model}. For independent draws $Y_1, ..., Y_n$ from a continuous 
random variable with CDF $F$, PDF $f$, and QF $Q$, and with a set of sample 
quantiles calculated by (\ref{eq:quant_df}), theorem \ref{thm:qclt} holds. 
Note that in (\ref{eq:qclt}), 
$N_K(\boldsymbol{\mu}, \boldsymbol{\Sigma})$ refers to a 
multivariate 
normal distribution with $K$ dimensions where $\boldsymbol{\mu}$ is the mean
vector and $\boldsymbol{\Sigma}$ is the covariance matrix. 
In (\ref{eq:qclt_covariance}) the 
operator $a \wedge b$ is the minimum of $a$ and $b$, with order of operations
such that $a \wedge b - ab = (a \wedge b) - ab$.

\begin{theorem}{\emph{Sample quantile central limit theorem}}
\label{thm:qclt}

Given a vector of length $K$ of probabilities 
$\boldsymbol{p} = (p_1, ..., p_K)$ and the corresponding quantile vector 
$\hat{\boldsymbol{Q}}_n(\boldsymbol{p}) = (\hat{Q}_n(p_1), ..., \hat{Q}_n(p_K))$, 
if $F$ is absolutely continuous for all $y \in \mathcal{Y}$ and is strictly 
increasing, then

\begin{equation}
    \label{eq:qclt}
    \sqrt{n}(\hat{\boldsymbol{Q}}_n(\boldsymbol{p}) - 
    \boldsymbol{Q}(\boldsymbol{p})) \overset{D}{\rightarrow} 
    N_K(0, \boldsymbol{\mathcal{K}})
\end{equation}
where $\boldsymbol{\mathcal{K}}$ is a $K\times K$ with the $ij^{th}$ 
entry 

\begin{equation}
    \label{eq:qclt_covariance}
    \kappa_{ij} = \frac{p_i \wedge p_j - p_i p_j}{f(Q(p_i)) f(Q(p_j))}.
\end{equation}
\end{theorem}

Theorem \ref{thm:qclt} has been known since the mid 20th century 
\cite[]{cramer1951mathematical}, and a thorough though not unique proof is 
found in \cite{walker1968note}. The entries in the covariance matrix in 
(\ref{eq:qclt_covariance}) may equivalently be written as
\begin{equation}
    \rho_{ij} = [p_i \wedge p_j - p_i p_j]q(p_i)q(p_j)
    \label{eq:qgp_simp_cov}
\end{equation}
where $q$ is the QDF. Another CLT in (\ref{eq:tqclt}) for a set of quantiles 
transformed by the underlying CDF $F(\hat{\boldsymbol{Q}}_n(\boldsymbol{p})) = 
(F(\hat{Q}_n(p_1)), ..., F(\hat{Q}_n(p_k)))$ is in (\ref{eq:qclt}) and can be 
derived from (\ref{eq:qclt}) by using the PIT and the Delta method 
\cite[]{parzen2004quantile}.

\begin{equation}
    \label{eq:tqclt}
    \sqrt{n}(F(\hat{\boldsymbol{Q}}_n(\boldsymbol{p})) - \boldsymbol{p}) 
    \overset{D}{\rightarrow} N_K(0, \boldsymbol{\Gamma})
\end{equation}
Here the covariance matrix $\Gamma$ has entries $\Gamma_{ij} = p_i \wedge p_j - 
p_i p_j$, making the asymptotic distribution in (\ref{eq:tqclt}) a 
Brownian bridge \cite[]{chow2009brownian}. This second result makes QM using 
the QGP model possible where $Q(p)$ is difficult to solve as is shown in the 
following section.

\section{Quantile Gaussian process model} \label{sec:qgp_model}

We consider the situation where one is given a set of data including a vector 
of $K$ probabilities $\boldsymbol{p} = (p_1, ..., p_K)$ and a vector of 
estimated quantiles at the given probability levels 
$\hat{\boldsymbol{Q}}_n(\boldsymbol{p}) = (\hat{Q}_n(p_1), ..., \hat{Q}_n(p_K))$. 
The set of quantiles may provide useful information about a distribution, but 
the information is limited and one may desire a more complete distribution.
The QGP model for QM and estimating a continuous distribution based on the CLT 
in (\ref{eq:qclt}) is in (\ref{eq:qgp}).

\begin{equation}
    \label{eq:qgp}
\hat{\boldsymbol{Q}}_n(\boldsymbol{p}) \sim N_K(\boldsymbol{Q}_{\theta}(\boldsymbol{p}), n^{-1} \boldsymbol{\mathcal{K}}_{\theta})
\end{equation}
Here $\theta$ is an unknown parameter vector to be estimated from the data. The 
covariance matrix $\boldsymbol{\mathcal{K}}_{\theta}$ has entries from the 
covariance function $\kappa_{\theta}(\cdot, \cdot)$ where 
\[
\kappa_{\theta}(p, p') = \frac{p\wedge p' - p p'}{f_{\theta}(Q_{\theta}(p)) f_{\theta}(Q_{\theta}(p'))}, \quad p, p' \in (0,1)
\]

The covariance function $\kappa_{\theta} (\cdot, \cdot)$ initially appears as 
if it will be difficult to evaluate and indeed can be, depending on the 
functional forms of $f_{\theta}$ and $Q_{\theta}$. But where these functions 
belong to distributions in a location-scale family, 
the covariance function can be greatly 
simplified. 
Noting that the covariance is a function of $f_{\theta}(Q_{\theta}(p)) = q_{\theta}(p)$, the QDF, if $\theta = (\mu, \sigma)$ where $\mu$ is a location parameter and $\sigma$ is a scale parameter, then $q_{\theta}(p) = \sigma q(p)$ where $q(p)$ is the QDF of a standard distribution having location 0 and shape 1. \cite{staudte2017shapes} calls this the location invariant and scale equivariant property. For location-scale distribution families, this greatly simplifies the form of $\kappa_{\theta}(\cdot,\cdot)$ so that it becomes
\[
    \kappa_{\theta}(p, p') = \sigma^{2}[p\wedge p' - p p'] q(p)q(p') 
\] 
Thus for estimating $\theta$ one only needs to estimate $\sigma^2$. A special 
case of the QGP for QM of a location-scale family is the normal QGP which we 
define and explore below.

\subsection{Normal QGP}

If the quantiles in a dataset 
$(\hat{\boldsymbol{Q}}_n(\boldsymbol{p}),\boldsymbol{p})$ are assumed to be 
calculated from a normal distribution $N(\mu, \sigma^2)$, and the desire is to 
estimate the parameters $\mu$ and $\sigma$ by modeling the quantiles according 
to the QGP in (\ref{eq:qgp}), then one may use the model (\ref{eq:norm_qgp}). 

\begin{equation}
    \label{eq:norm_qgp}
\hat{\boldsymbol{Q}}_n(\boldsymbol{p}) \sim N_K 
\left( \mu + \sigma \boldsymbol{\Phi}^{-1}(\boldsymbol{p}),\frac{\sigma^2}{n} 
\boldsymbol{\Psi} \right)
\end{equation}
Here $\boldsymbol{\Phi}^{-1}(\boldsymbol{p}) = (\Phi^{-1}(p_1), ..., 
\Phi^{-1}(p_K))$ is a known vector since $\boldsymbol{p}$ is given, and 
$\boldsymbol{\Psi}$ is a known $K\times K$ matrix with $ij^{th}$ entry

\[
    \Psi_{ij} = \frac{2 \pi (p_i\wedge p_j - p_i p_j)}{\text{exp}\{-\frac{1}{2}[\Phi^{-1}(p_i)^2 + \Phi^{-1}(p_j)^2]\}}.
\]
Model (\ref{eq:norm_qgp}) can then be viewed as an atypical normal linear 
regression model. Two aspects that make this model atypical relative to the 
standard linear regression model are that the slope parameter $\sigma$ must be 
greater than 0 and that $\sigma$ is both a regression coefficient and part of 
the variance.
Note also that the sample size $n$ is included as part of the variance. If $n$ 
is known, then it may be multiplied through $\Psi$ and forgotten. If unknown, 
it can then be accounted for as an unknown part of the variance. 

Take for instance the data 
$\boldsymbol{X} = \left(\textbf{1}, \Phi^{-1}(\boldsymbol{p})\right)$ and the 
parameter vector $\beta = (\mu, \sigma)$. Assuming the sample size $n$ is 
unknown and setting $\sigma^2/n = \gamma^2$, model (\ref{eq:norm_qgp}) then 
becomes
\[
    \hat{\boldsymbol{Q}}_n(\boldsymbol{p}) \sim N_K\left(X\beta, 
\gamma^2 \boldsymbol{\Psi}\right)
\]
In this model, one simplifying assumption may be to treat $\sigma$ and 
$\gamma^2$ separately and proceed to fit a standard linear regression model. 
All frequentist and Bayesian results of the linear regression model apply, 
including the existence of conditionally conjugate prior distributions. The 
positive constraint on $\sigma$ can also be dealt with without adding much 
complication (see \cite{gelman2013bayesian} pgs. 377-378). 
If, for a certain problem it is important to make inference on the unknown $n$, or $\sigma$ and 
$\gamma^2$ cannot treated as separate, then one may estimate the two parameters 
by assigning appropriate prior distributions to $\sigma$ and $n$ and reverting 
back to (\ref{eq:norm_qgp}). In section \ref{sec:simulation_analyses}, we 
analyze a normal QGP in a simulation study where we assign independent prior 
distributions to $\sigma$ and $n$, and $\sigma$ is treated as both a regression 
coefficient and as part of the variance.

For any QGP model where the QF belongs to a location-scale family, the 
discussed relation to the normal linear regression model applies. The only 
difference in modeling becomes formulating the matrix $\Psi$ using the proper 
QDF, and the data $\boldsymbol{X}$ with the proper quantile function. Many 
situations will require a more complicated QF which doesn't allow for modeling 
the data as a linear regression. Below we consider for instance a finite 
mixture distribution. 

\subsection{Finite normal mixture QGP}

For a continuous random variable distributed according to a finite mixture 
distribution, the CDF takes the form of (\ref{eq:norm_mix}). Here there are $C$ 
component distributions where $c \in \{1, ..., C\}$, $F_{\theta_c}$ is a 
continuous CDF with parameter $\theta_c$, and $w_c > 0$ is a weight such that 
$\sum_{c = 1}^C w_c = 1$. 

\begin{equation}
    \label{eq:norm_mix}
    F_{\theta}(x) = \sum_{c = 1}^C w_c F_{\theta_c}(x)
\end{equation}

Often $F_{\theta}$ will not be easily invertible, thus $Q_{\theta}$ will be a 
complicated function which can be evaluated only via numerical optimization 
which is often computationally expensive and/or inaccurate. A simple solution 
is to model the PIT transformed quantiles from the data rather than modeling 
the quantiles directly. The QGP model (\ref{eq:qgp}) is then reformulated to be 
model (\ref{eq:pit_qgp}), where  $\Gamma_{ij} = p_i\wedge p_j - p_i p_j$ as in
(\ref{eq:tqclt}).   

\begin{equation}
    \label{eq:pit_qgp}
F_{\theta}(\hat{\boldsymbol{Q}}_n(\boldsymbol{p})) \sim 
N_K(\boldsymbol{p}, n^{-1} \boldsymbol{\Gamma})
\end{equation}

By modeling the PIT quantiles, the only function which requires evaluating is 
the CDF $F_{\theta}$. Here both $\boldsymbol{p}$ and $\boldsymbol{\Gamma}$ are 
given, and solving for the transformation 
$F_{\theta}(\hat{\boldsymbol{Q}}_n(\boldsymbol{p}))$ is unnecessary for our 
purposes. A Bayesian fit of (\ref{eq:pit_qgp}) is straightforward, requiring 
only that $F_{\theta}(\hat{\boldsymbol{Q}}_n(\boldsymbol{p}))$ be evaluated as 
part of the acceptance ratio of a Markov chain Monte Carlo (MCMC) iteration. 

In section \ref{sec:simulation_analyses}, we analyze PIT transformed QGP model 
(\ref{eq:pit_qgp}) where $F_{\theta}$ is set as a normal mixture distribution. 
Fitting this model proved faster 
and produced better results than trying to fit the QGP model (\ref{eq:qgp}) and 
evaluating $Q_{\theta}$ the QF of a normal mixture. In fact we found that in 
cases where $F_{\theta}$ was a simpler function, such as a normal or 
exponential distribution, fitting model (\ref{eq:pit_qgp}) was at least as fast 
and the results were at least as good as fitting (\ref{eq:qgp}). Thus when 
possible, we elected to fit (\ref{eq:pit_qgp}).

\subsection{Competing quantile matching methods}

A number of methods already exist for QM. Here we briefly review four of those 
methods, and in section \ref{sec:simulation_analyses} we compare performance 
between these and the QGP. The four methods include spline interpolation (SPL), 
kernel density estimation (KDE), an order statistics based model (ORD), and an 
independent quantile model (IND). The first two of these methods are 
non-parametric methods and neither of which include modeling the uncertainty 
of the quantiles.

The SPL method was used by \cite{gerding2023evaluating} and 
\cite{shandross2024hubensembles} in order to estimate a CDF function given 
quantile forecasts from disease outbreak forecast hubs. SPL is an interpolation 
where monotonic cubic splines are fit to pass through each given quantile and 
predict a function for all values between given quantiles and beyond the 
extreme values. An \texttt{R} package used for fitting SPL and which we use in 
section \ref{sec:simulation_analyses} is \texttt{distfromq} 
\cite[]{ray2024quantmatch}.

The KDE method treats given quantiles as if they constitute a random sample 
from a distribution and applies kernel smoothing to estimate a density 
function. KDE requires selecting a kernel function, often a Gaussian kernel is 
chosen, and applying that function to each draw of a sample. 
\cite{gyamerah2020probabilistic} applied KDE smoothing with an Epanechnikov 
kernel to quantiles estimated via three different machine learning methods 
used for predicting crop yield. They then combined the estimated densities 
into an ensemble prediction. \cite{he2016short} also use KDE to estimate 
density forecasts from quantiles to produce energy forecasts. We use the 
\texttt{evmix} package for performing QM using KDE \cite[]{yang2018kde}.

One of the more common methods for QM is to select a CDF or QF function and 
then select model parameters which give a least squares fit to the estimated 
quantiles as done by \cite{li2019combining}. 
The \texttt{R} package \texttt{rriskDistributions} performs this least squares 
fit for some standard distributions \cite[]{belgorodski2017quantilemse}. 
Independent random error may be included to the best fit QF allowing for 
estimation of the variability of the quantiles \cite[]{nirwan2020bayesian}. 
In section \ref{sec:simulation_analyses}, we analyze model 
(\ref{eq:ind_quantile}) as a proxy for the IND model to compare with 
other methods. 
\begin{equation}
    \label{eq:ind_quantile}
    F_{\theta}(\hat{Q}(p_k)) \overset{ind}{\sim} N(p_k, \sigma_{\rho}^2)
\end{equation}
Here $\hat{Q}_k$ is the estimated sample quantile at probability $p_k$. 
The major difference between model (\ref{eq:ind_quantile}) and model 
(\ref{eq:pit_qgp}) is that the error for each quantile in a set of quantiles is 
considered independent thus ignoring the effects of sample size and correlation 
suggested by the CLT in (\ref{eq:tqclt}). Note that the parameter 
$\sigma_{\rho}$ is not a part of $\theta$, the parameter of the ``true`` 
distribution estimated by QM but instead is meant to capture independent 
error among the sample quantiles.

The final method we include is the ORD model introduced by 
\cite{nirwan2020bayesian}. This model relies on the definition of sample 
quantiles being order statistics. In the ORD model, the joint distribution of a 
set of order statistics is the model likelihood. The likelihood of a set of 
order statistics from a continuous distribution is a function of both the CDF 
and PDF of a continuous distribution. ORD is the method most similar to QGP and 
it provides exact inference of quantile uncertainty whereas the QGP provides 
asymptotic inference of uncertainty. 

For assessing QM, there are two aspects we analyze. The first is a model's 
ability to estimate parameters and the second is how far away a fit QM 
distribution is from a true distribution. Of the methods previously listed, 
parameter inference is only possible for the QGP, IND, and ORD models. These 
are also the only methods which provide uncertainty quantification for quantile 
values. To analyze how far an estimated distribution is from a true 
distribution, a distance measure must be utilized. We outline three of these 
distances below.

\subsection{Distance measures}

A number of metrics for measuring the distance between two continuous 
univariate distributions exist. Those we consider in this manuscript are the 
Wasserstein distance (WD) and a slight modification of it, the total variation 
(TV) or the statistical difference, and the Kullback-Leibler divergence (KLD). 
A brief overview of these and other metrics, some of their statistical uses 
and properties, and relationships between metrics is found in 
\cite{gibbs2002choosing}. These metrics were used in section 
\ref{sec:simulation_analyses} to assess how well different QM methods 
approximate a true distribution.

The WD is used to measure the distance between two univariate random variables 
by their CDFs or QFs and has many applications in mathematics, optimization, 
and statistics, see \cite{panaretos2019statistical} for a review of the WD and 
its use in statistics. The $p$-WD is defined for two continuous random 
variables $X$ and $Y$ with respective CDFs $F_X$ and $F_Y$. In terms of the 
two CDFs the $p$-WD is defined in (\ref{eq:wass_dist_cdf}), and the $p$-WD in 
terms of the corresponding QFs is defined in (\ref{eq:wass_dist_quant}). 

\begin{equation}
    \label{eq:wass_dist_cdf}
    WD_p (F_X, F_Y) = \left( \int_{\mathbb{R}} |F_X(t) - F_Y(t)|^p dt\right)^{1/p}
\end{equation}

\begin{equation}
    \label{eq:wass_dist_quant}
    WD_p (F_X, F_Y) = \left( \int_0^1 |F^{-1}_X(t) - F^{-1}_Y(t)|^p dt \right)^{1/p}
\end{equation}

We define a modified version of the WD, the uniform 1-WD (UWD1) in 
(\ref{eq:pit_wass_dist}). Here we take $F_X$ to be the CDF of a continuous 
random variable $X$. For a different continuous random variable $\xi$ with CDF 
$F_{\xi}$, $F_X(\xi)$ is a random variable with support on $[0,1]$. We let 
$F_{X,\xi}$ be the CDF of the random variable $F_X(\xi)$. Then the measure in 
(\ref{eq:pit_wass_dist}) measures how close $F_X(\xi)$ is the   CDF of 
a standard uniform distribution with CDF $\mathcal{U}$, 
noting that by the PIT, if 
$\xi \overset{D}{=} X$, $F_X(\xi)$ is the random variable of a standard uniform 
distribution. The integral in (\ref{eq:pit_wass_dist}) takes values between
0 and 1/2 \cite[]{645854}, and multiplying by two 
ensures the measure takes values between 0 and 1. This makes the UWD1 more
interpretable than the WD
where a value near 0 suggests $\xi$ and $X$ are ``near'' each other, and a value
near 1 suggests they are far apart.   

\begin{equation}
    \label{eq:pit_wass_dist}
    UWD_1(F_{X,\xi}, \mathcal{U}) = 2\int_0^1 |F_{X,\xi} (u) - x|du
\end{equation}


The TV, defined in (\ref{eq:tot_var}), is a distance between distributions 
measured in terms of the PDFs $f$ and $g$. The TV takes values between 0 and 1, 
and it is closely related to the distance used in 
\cite{sgouropoulos2015matching} to measure the goodness of QM.  

\begin{equation}
    \label{eq:tot_var}
    TV(f,g) = \frac{1}{2} \int_{-\infty}^{\infty}|f(x) - g(x) | dx
\end{equation}


The final metric between distributions we consider is the KLD defined in 
(\ref{eq:kld}). The KLD is not a true metric but has many useful properties 
and applications. It may be interpreted as the expected divergence if one is 
using $f$ to approximate $g$.

\begin{equation}
    \label{eq:kld}
    KLD(g || f) = \int_{\mathcal{X}} 
    g(x) \text{log} \left( \frac{g(x)}{f(x)} \right) dx
\end{equation}

\section{Quantile matching methods comparison on simulated data} \label{sec:simulation_analyses}

In this section we present simulation studies assessing the QM of the
QGP model and compare results with those of the SPL, KDE, IND, and ORD matching
methods. The QGP, IND, and ORD models were each fit via Hamiltonian Monte Carlo
(HMC) sampling using the \texttt{Stan} software and the \texttt{cmdstanr}
package developed and maintained by the \cite{stan2024manual}
\cite[]{gabry2022stan}. Simulation studies are done to assess parameter
estimation and inference as well as QM estimation of a distribution.

\subsection{Parameter estimation and quantile matching for known distribution families} 

Where the distribution family is known, one goal of QM may be to estimate and 
make inference on model parameters. To that end, we analyzed 
parameter estimation and inference for normal and exponential family 
distributions with known parameters. We analyzed the QGP model from 
(\ref{eq:pit_qgp}), the IND model, and the ORD model. We note that model 
(\ref{eq:pit_qgp}) allows one to do inference for both the model parameter 
$\theta$ and for an unknown sample size $n$. The ORD model also allows for 
estimation of an unknown $n$. For the QGP and ORD models, we include modeling 
cases where $n$ is known and where $n$ is unknown and estimated. We denote the 
models where $n$ is known as QGP-n and ORD-n. We also compared the distance 
between the true model and the predictive distributions estimated via QGP, IND, 
ORD, SPL, and KDE, and we include a comparison of the QGP and ORD models 
where the true distribution is a quantile defined distribution.

\subsubsection{Normal parameter estimation}
In this simulation study, we fit the ORD and IND models to quantiles estimated 
from independent samples from a normal distribution and compare parameter 
estimation with fits from the QGP model. Quantiles were simulated by first 
drawing a sample of size $n$ from a known distribution then estimating $K$ 
quantiles given probabilities $\{p_1, ..., p_K\}$. The ORD, IND, and QGP models 
were then fit to the quantiles where $F$ was set to be the CDF of a normal 
distribution. For each of 
$n \in \{50, 150, 500, 1{,}000, 5{,}000\}$ and 
$K = 23$, there 
were 500 simulation replicates. The data were simulated from a normal 
distribution with mean $\mu = 4$ and standard deviation $\sigma = 3.5$.
For each of QGP, ORD, and IND models, we assigned the same prior distributions 
to $\mu$ and $\sigma$. When $n$ is unknown, we also assigned it a prior 
distribution. 
As one will generally have access to the quantiles being fit, 
it is reasonable that an
informative prior, centered near where the data is
centered, be assigned to $\mu$. 
We took this approach herein, and for scale parameters,
we assigned normal priors truncated at 0 as recommended by
\cite{gelman2006prior}.
The prior distributions were $\mu \sim N(5,7^2)$,
$\sigma \sim N(0, 6^2) \mathds{1} \{\sigma > 0\}$, and
$n \sim N(0,3000^2)\mathds{1}\{n > 0\}$.
For the IND model the prior on the independent parameter 
$\sigma_{\rho}$ from model (\ref{eq:ind_quantile}) is 
$1/\sigma_{\rho} \sim N(0, 3000^2)\mathds{1}\{\sigma_{\rho} > 0\}$,
which is similar to the prior on $n$ for 
the QGP and ORD models. For each fit, the HMC chain was run for 60,000 draws 
with the first 10,000 discarded as a burn-in.

\begin{figure}[hbt!]
  \centering
  \includegraphics[width=1\linewidth]{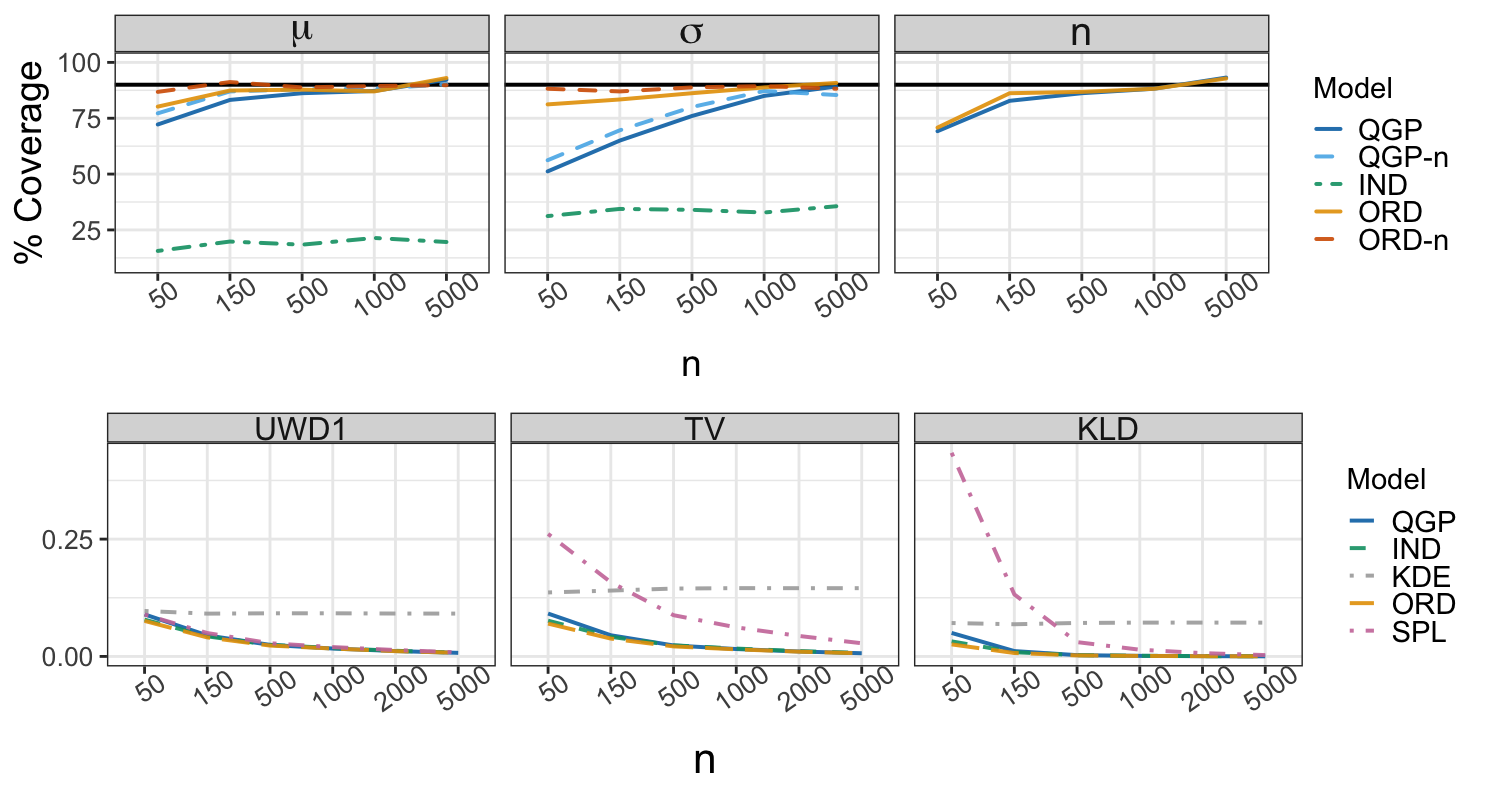}
\caption{Posterior coverage (top) for models of $K = 23$ quantiles
calculated as the percentage of times the 
true parameter fell within the modeled 90\% credible interval over the 500 
replications. Coverage is faceted by the normal parameters $\mu$, $\sigma$,
and shown by increasing sample size 
($x$-axis). The five models QGP, ORD, QGP-n, ORD-n, and IND are colored as 
shown the legend. The horizontal line (black) is at the nominal 90\% level. 
Only QGP and ORD appear for inference for the parameter $n$ as they are the 
only two models which 
estimate an unknown $n$.
Distance between the true distribution and the estimated QM predictive 
distribution (bottom) averaged over the 500 replications. Distances include the 
UWD1, TV, and KLD for increasing sample 
size ($x$-axis).}
\label{fig:normal_cov_dists}
\end{figure}

The top of figure \ref{fig:normal_cov_dists} shows the 90\% coverage of 
the posterior distribution credible intervals as $n$ increases for 
$\mu$, $\sigma$ and $n$ for five models. QGP-n and ORD-n models where $n$ is 
known are included along with QGP, ORD, and IND. 
The 90\% credible intervals of the 
parameters were calculated by computing the $0.05^{th}$ and $0.95^{th}$ 
quantiles from the posterior distribution samples. The coverage percentage is 
calculated as the percentage of the 500 replicates for 
which the true parameter 
value was within the 90\% credible interval.
As expected, the nominal coverage for the QGP and ORD models is better than 
that of the IND model, and for the two models where $n$ is known the coverage 
is better than where it is unknown, particularly for lower values of 
$n$ and $K$. The coverage for the ORD models appears to be slightly better 
than for the QGP models though not by much, and the difference
decreases as $n$ increases. 
The superior performance of the ORD model is possibly because it
provides exact inference 
whereas the QGP model provides asymptotic inference.

To analyze how well QM methods produced predictive distributions which 
approximate the true distribution $N(4, 3.5^2)$, we measured the distance 
between the estimated predictive distributions of QGP, ORD, and IND fits to 
the true distribution. 
We also measured the distances of the SPL and KDE QM fits to the true 
distribution. The distances were measured via the UWD1, TV, and KLD metrics. 
The UWD1 for each of QGP, ORD, QGP-n, ORD-n, and IND were calculated as 
follows. For $\{\theta_m\}^M$ being $M$ draws from the posterior distribution
of $\theta$ where $\theta_m = (\mu_m, \sigma_m)$, the QM posterior predictive 
distribution was simulated by repeatedly sampling a $\theta_m^*$ from 
$\{\theta_m\}^M$, then sampling 
$X_m$ from a distribution with CDF $F_{\theta_m^*}$. Repeating this for 
$M = 50,000$ times gives a QM posterior predictive sample. For the CDF 
$F_{\theta}$ where $\theta$ is the true parameter, $\xi_m = F_{\theta}(X_m)$ 
was calculated and the empirical CDF $\hat{F}_{\xi}$ was calculated from the 
sample $\{\xi_m\}^M$. $F_{\xi}$ in (\ref{eq:pit_wass_dist}) was then replaced 
by $\hat{F}_{\xi}$ to calculate UWD1. For SPL and KDE $F_{\xi}(x)$ was replaced 
by the respective estimated CDFs. The \texttt{integrate} function in \texttt{R} 
was used for calculating the UWD1 in (\ref{eq:pit_wass_dist}).

To estimate TV and KLD, rather than sampling from the posterior distribution, 
the marginal means for the parameters were calculated from the posterior 
distribution samples. These marginal means  are 
$\Bar{\theta}_M = (\Bar{\mu}_M, \Bar{\sigma}_M)$ where $\Bar{\mu}_M$ and 
$\Bar{\sigma}_M$. To calculate TV, 
$f$ in (\ref{eq:tot_var}) is replaced by $f_{\Bar{\theta}_M}$ and $g$ is 
replaced by $f_{\theta}$. The TV calculation was done using the 
\texttt{integrate} function in \texttt{R}. The KLD was estimated also using 
$f_{\Bar{\theta}_M}$ and $f_{\theta}$, but was done via Monte Carlo sampling. 
That is (\ref{eq:kld}) was estimated by

\[
    \widehat{KLD}(f_{\theta} || f_{\Bar{\theta}_M}) = 
    \frac{1}{S}\sum_{s = 1}^S \text{log} 
    \left(\frac{f_{\theta}(Y_s)}{f_{\Bar{\theta}_M}(Y_s)}\right)
\]
where $S$ is the total number of Monte Carlo samples and
$Y_s \overset{iid}{\sim} F_{\theta}$. For both SPL and KDE, 
$f_{\Bar{\theta}_M}$ is the estimated PDF.

The bottom of figure \ref{fig:normal_cov_dists} shows the UWD1, TV, and 
KLD metrics averaged over the 500 simulation replicates as $n$ increases for 
the five QM methods. The QGP, ORD, and IND are nearly indistinguishable and 
outperform the SPL and KDE in almost every case. 
For UWD1, the SPL performs similarly to the parametric QM methods, but for the 
TV and KLD, the SPL performs much worse, especially for smaller sample sizes. 

This study shows the ability of the QGP QM model to estimate and perform 
inference on model parameters where the true distribution family is known. It 
also shows the QGP's ability to produce a posterior predictive distribution 
which closely matches a true distribution, according to several metrics, 
relative to other QM methods. Between parameter inference and predicting the 
true distribution, QGP is superior to all methods we compared it with except 
for the ORD model which has similar results to --or for small sample size
slighty better than-- the QGP. However, we feel the QGP's linear model form is
easier to understand than the ORD model and thus preferable for interpretation.
Additional plots for this study and for a similar simulation 
study where the data was simulated from an exponential family distribution are 
in the supplementary materials.
The results of the exponential family study are similar to those of the 
normal distribution family study.

When fitting the Bayesian models both in the normal and in the exponential 
setting, the IND model tended to fit the fastest, followed by the ORD models, 
and finally the QGP models. The time to fit however was very fast, usually no 
more than seven seconds except for the case where $K = 50$ quantiles, in which 
case the QGP model where $n$ was unknown often took around 20 seconds to fit. 
However, with $K= 50$ the QGP-n model continued to fit very rapidly, fitting in 
under seven seconds.

The analysis above show the usefulness in using the QGP model both in parameter 
estimation and in QM when given sample quantiles. In the normal case, the QGP 
showed improved parameter inference over the IND model and allows for making 
inference on the sample size $n$. Compared to the SPL and KDE QM methods, the 
QGP fits tend to be closer in distance to the true distribution. The QGP model, 
however, performs about the same or for smaller $n$ slightly poorer than the
ORD model in terms of inference and QM. However, we feel the QGP is a simpler
model to
understand than the ORD and easier to interpret.
The following section outlines a situation where the QGP model may be a better 
option for QM than the ORD model.

\subsubsection{Quantile defined distributions}
The normal and exponential families both have CDF and PDF functions which are 
either available in closed form or are easy to evaluate with software which is 
widely accessible. The next study we performed was done to compare the QGP and 
ORD models where the distribution family was a quantile defined distribution 
which lacks a CDF that is easy to evaluate. The purpose of this study was to 
show an example where one may prefer to perform QM using the QGP rather than 
the ORD.

\begin{figure}[hbt!]
    \centering
    \includegraphics[scale=.22]{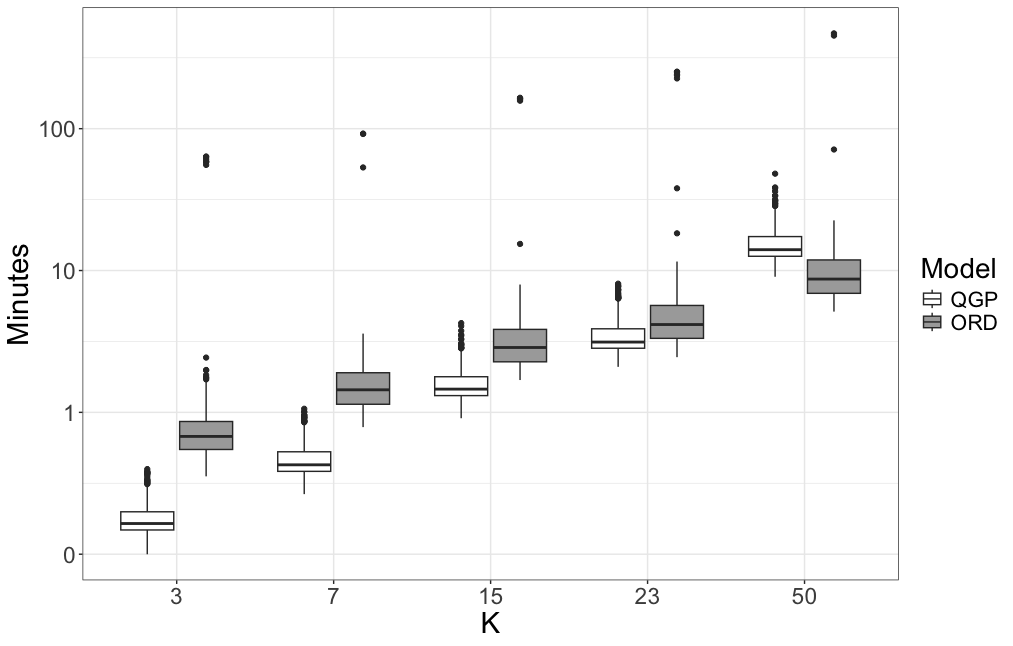}
    \caption{Boxplots of time to fit a model in seconds ($y$-axis) for the 500 
    replicates in the simulation study for QM of the generalized lambda 
    distribution (GLD) for $K \in \{3, 7, 15, 23, 50\}$ ($x$-axis). Boxplots 
    are separated by QM methods QGP (red) and ORD (blue). The $y$-axis is on 
    the $\text{log}_{10}$ scale.}
    \label{fig:tuk_time}
\end{figure}

The GLD is reviewed in section \ref{sec:quant_def_dist}, and a simulation 
carried out similarly to the previous studies was performed with GLD as the 
true distribution. However, QM was only done by the QGP and ORD models. 
Because of the relative difficulty of evaluating the CDF of the GLD and the 
ease of evaluating the QF, the CDF transformed QGP model in (\ref{eq:qgp}) 
instead of (\ref{eq:pit_qgp}) was used. For the ORD model, evaluating the CDF 
was required for modeling. Evaluating the CDF of the GLD requires using an 
algebraic solver, and it took longer to code the ORD model in \texttt{Stan} 
due to a lack of readily available software which we were aware of. 
Once working, however, both QGP and ORD models fit the estimated quantiles 
well, however, the ORD usually required more time to fit than the QGP. 
Figure \ref{fig:tuk_time} shows boxplots for the 500 replicates for different 
values of $K$ quantiles. In all cases except for $K=50$ the QGP tends to fit 
much faster than ORD. Note that the $y$-axis is on the logarithmic scale. Even 
at $K=50$ some outliers for the ORD model required hundreds of minutes to fit 
whereas the longest required time for fitting a QGP model was a few dozen
minutes.

 The MLD was also reviewed in section \ref{sec:quant_def_dist}. We were able 
 to fit a QGP model for the MLD which appeared to fit sample quantiles well, 
 but we were unable to fit a working ORD model. The algebraic solver struggled 
 to evaluate the CDF, and posterior sampling in \texttt{Stan} was extremely 
 slow. When fitting finally finished, the result was a very poor fit. Perhaps 
 with more time and effort we could have made a working model, but the QGP 
 worked well enough that we did not feel it was worth the effort.

\subsection{Matching unknown distributions} \label{sec:uk_matching}

The simulation studies in this section are for the situation where the true 
distribution family is unknown, unimportant, or too complex to be practically 
evaluated but where one still wishes to recover or approximate the 
distribution. We selected three distributions each with a different shape to 
analyze how well the distributions can be approximated by QM using the QGP and 
the competing methods. 

With the true distribution family being unknown, we made the modeling decision 
to set $F$ in the QGP model of (\ref{eq:pit_qgp}) to be a mixture of normals 
distribution taking the form of (\ref{eq:norm_mix}). This decision to use
a normal mixture was 
based on the common claim that any continuous distribution may be approximated 
by a finite mixture of normal distributions 
\cite[]{peel2000finite,nguyen2019approximations,nguyen2020approximation}. This 
claim may be optimistic, but we found that with enough components,
the normal mixture 
distribution can approximate the non-normally shaped distributions reasonably 
well. However, any function which meets the technical definition of a CDF would 
be appropriate to use for $F$, see \cite{gasthaus2019probabilistic} for an 
example.
After a small analysis comparing fits of one to six normal components,
the number of component 
distributions was chosen to be $C = 4$. The results, included in the 
supplementary materials, show that as the number of components in the mixture
distrubtion increases, the distribution approximation improves. But for the 
cases considered, the improvement from four to five components or more is small.

The three distributions from which data were simulated and sample quantiles 
estimated and to which QM methods were fit were the extreme value (EV) 
distribution with location 0 and scale 1 $EV(0,1)$, Laplace (La) with location 
0 and scale 1 $La(0,1)$, and a two component normal mixture (MIX) 
$w N(-1, 0.9) + (1-w)N(1.2, 0.6)$ where $w = 0.35$. From the three 
distributions, 500 replicates of $K = 23$ quantiles were simulated for each of 
$n \in \{50, 150, 500, 1{,}000, 2{,}000, 5{,}000\}$. QM for each replicate of 
simulated quantiles was done using QGP, ORD, IND, SPL, and KDE. For the QGP, 
ORD, and IND models, $\theta = \{\theta_1, \theta_2, \theta_3, \theta_4\}$ 
where $\theta_c = (\mu_c, \sigma_c)$ for $c \in \{1,2,3,4\}$. The prior 
distributions assigned were $\mu_c \overset{ind}{\sim} N(5, 7^2)$,
$\sigma_c \overset{ind}{\sim} N(0, 6^2)\mathds{1}\{\sigma_c > 0\}$,
$w \sim Dirichlet(\boldsymbol{1}_C)$, and
$\nu \sim N(0, 3000^2)\mathds{1}\{\nu > 0\}$
where $\nu = n$ for the QGP and ORD models 
and $\nu = 1/\sigma_{\rho}$ for the IND model. 
Often when implementing posterior sampling for a mixture distribution model, 
one deals with an issue called the label-switching problem. This is where for a
mixture distribution parameter $\theta = \{\theta_1,...,\theta_C\}$, the model 
likelihood is the same for different permutations of $\theta$. This lack of 
identifiability for elements of $\theta$ makes parameter inference useless, 
but the predictive distribution may still be close to the true distribution 
\cite[]{stephens2000dealing}.
Because of this, the HMC posterior sampling chain was run longer than in 
studies of the previous section. The chain was run for 80,000 steps where the 
first 20,000 draws were discarded as a burn-in. For this study, we were 
concerned only with QM and not with parameter inference, so the fact that the 
parameters are unidentifiable was not considered critical. However, when 
assessing the TV and KLD between the QM and true distributions, variational 
Bayes (VB) in \texttt{Stan} was used instead of HMC for fitting the models 
\cite[]{kucukelbir2015automatic}. This decision was made because of the lack of 
parameter identifiability and because the marginal parameter means are used to 
estimate the PDF. While MCMC methods perform better on parameter inference, 
VB methods have been shown to have predictive performance comparable to MCMC 
\cite[]{blei2017variational}.

\begin{figure}[hbt!]
\centering
  \centering
  \makebox[\textwidth][c]{%
        \adjustbox{trim=0 0 0 0,clip}{\includegraphics[width=1.3\textwidth]{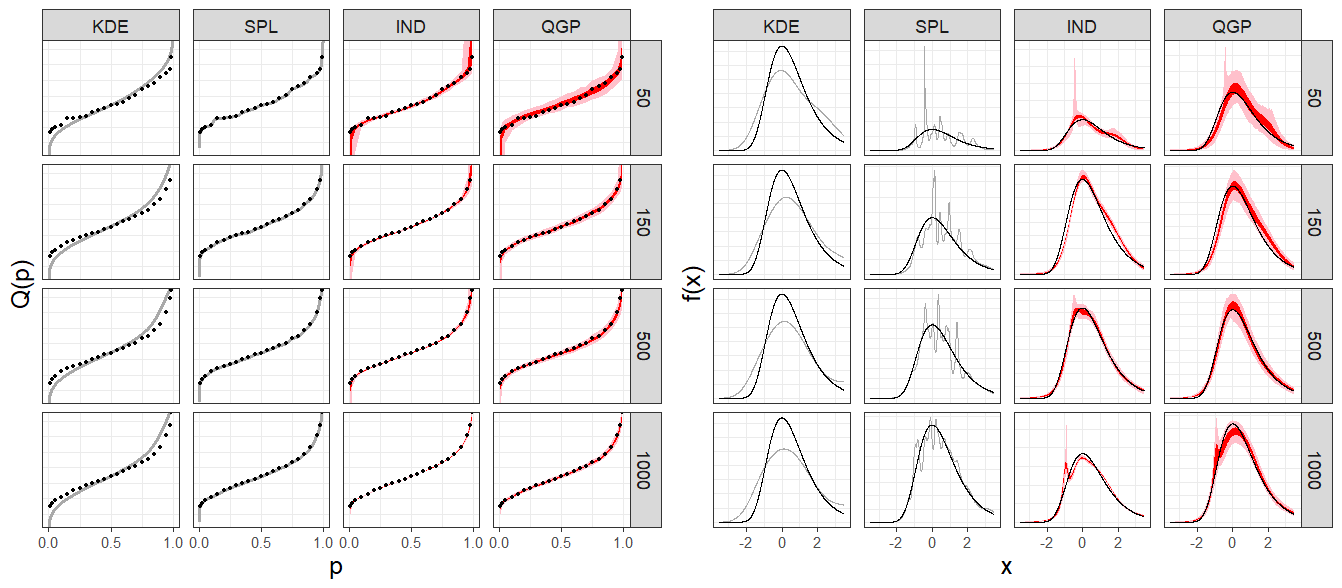}}%
    }

\caption{QM fits of $K=23$ quantiles by KDE, SPL, IND, and QGP for 
$n \in \{50, 150, 500, 1{,}000\}$. The quantiles were sampled from the extreme 
value distribution $Ev(0,1)$. The quantile fits (left) show the true quantiles 
(black) with either the QM fit line (grey) or the credible intervals of 50\% 
(red) and 95\% (pink). 
The estimated PDF plots (right) show the true PDF (black) with either a the QM 
estimated PDF (grey) or the credible intervals of 50\% (red) and 95\% (pink).}
\label{fig:evd_fits}
\end{figure}

Figure \ref{fig:evd_fits}, 
shows examples of fits of $K=23$ quantiles simulated from the EV distribution
for different sample sizes $n$. Included QM methods 
are KDE, SPL, IND, and QGP. ORD was excluded to make visualization easier, but 
the fits are very similar to the QGP fits. The KDE and SPL fits provide no 
uncertainty estimation of the quantiles, and return only a continuous function. 
The SPL fits show a lot of wiggle with the wiggle decreasing as $n$ increases, 
whereas the KDE fits do not show as much wiggle, but the KDE fits the 
distribution tails poorly even as $n$ increases. The IND and QGP models provide 
uncertainty in the fits, but the QGP is much more conservative with wider 
intervals which provide superior coverage of the quantiles and the PDFs.

\begin{figure}[hbt!]
\centering
\includegraphics[width=.9\linewidth]{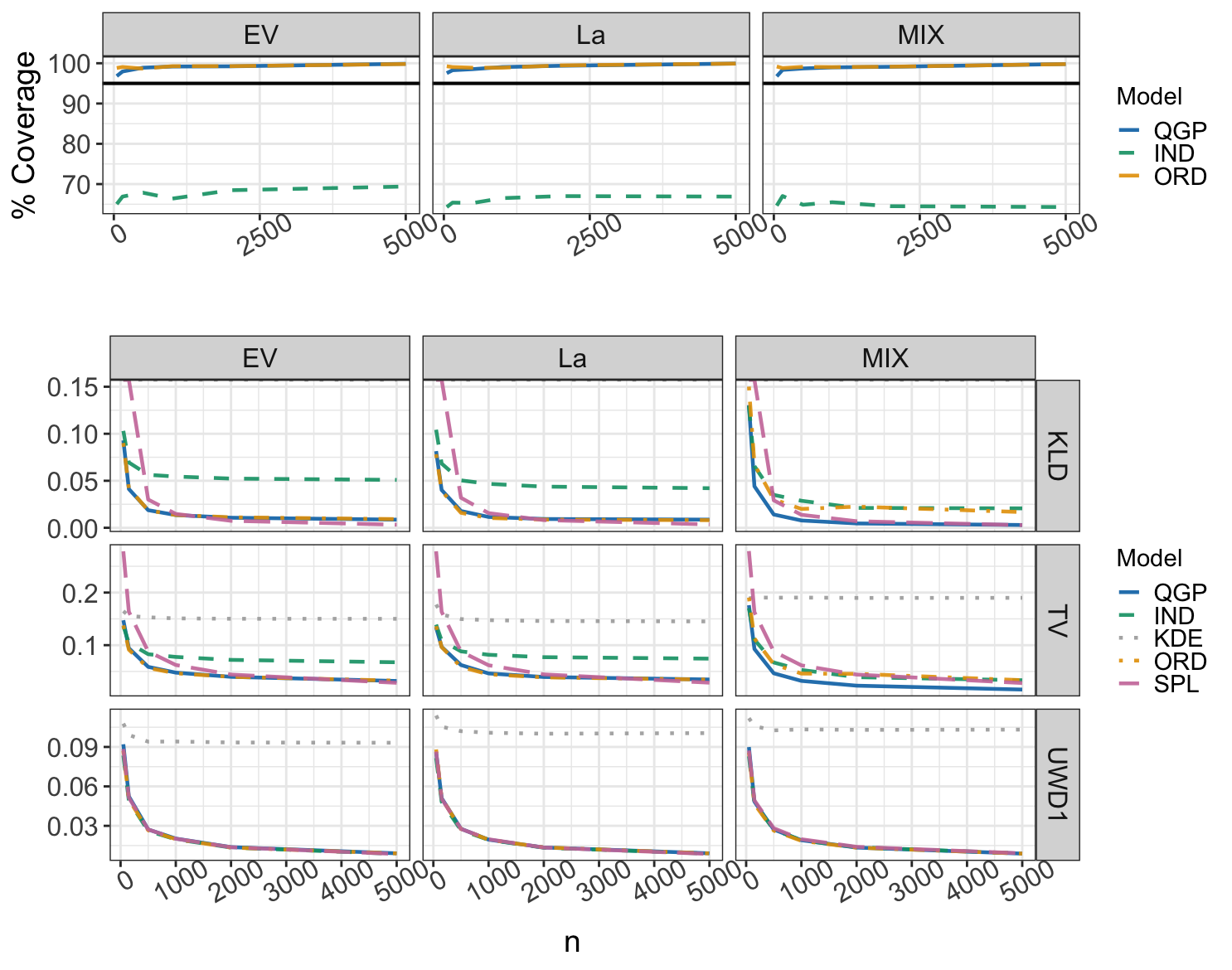}
\caption{
(top) Percent coverage of the true quantile values for select quantiles for 
500 simulated replicates of sample quantiles averaged over all $K = 23$
quantiles faceted by the true EV, La, and normal mixture distributions.
The plots shows percent coverage for the 95\% credible intervals. The models 
included are QGP, ORD, and IND. The vertical bar (black) shows the nominal 
coverage level.
(bottom) Distance between the true distribution and the QM estimated 
distribution 
averaged over 500 simulation replicates and measured by KLD, TV and UWD1. 
Plots are faceted by distribution with the distributions being 
EV, La, and MIX. 
Lines are colored and shaped by QM method, and are drawn by increasing sample 
size $n$.}
\label{fig:uk_cover}
\end{figure}

The top of figure \ref{fig:uk_cover} shows the 95\% simulated coverage of the 
true quantiles 
for QGP, ORD, and IND fits of quantiles from the EV, La, and MIX distributions. 
The percent coverage is 
averaged over all $K = 23$ quantiles and all 500 simulated replicates.  
The plots show that the QGP and ORD models are largely able to meet and exceed 
the nominal coverage where the IND model coverage is low. Because of the time it
took to evaluate the QF when calculating coverage, calculation was done on 
a thinned sample of the posterior distribution where only every $100^{th}$ HMC 
draw was kept so that the coverage was calculated using 6,000 posterior draws 
rather than the full 60,000.

The bottom of \ref{fig:uk_cover} shows the UWD1, TV, and KLD distances between 
QM and true distributions averaged over 500 simulation replicates for the 
five QM methods. In the UWD1 plot, there appears to be little or no difference
in performance between the QGP, ORD, IND, and SPL fits, but each of these
outperforms the KDE fits. For TV, the QGP, ORD, and SPL perform similarly, 
though the QGP does the best for the MIX distribution.
And for KLD, the QGP and ORD perform the best for lower values of $n$, but as
$n$ increases, SPL appears to perform the best. Additional plots and results
from these analyses are found in the supplementary materials.

%

This section shows the ability of the QGP model to accurately perform parameter 
inference and QM when given sample quantiles. Although the inference for QGP 
is asymptotic, the QGP performance in inference and in QM is similar to that 
of ORD which provides exact inference. There are also circumstances where the 
QGP may outperform the ORD as shown in cases where the QF is easily evaluated 
but where the CDF is difficult to evaluate. Another case where the QGP 
outperformed the ORD was in QM of the normal mixture distribution. An 
advantage that the QGP has over SPL and KDE is that it provides accurate 
uncertainty estimation of quantile uncertainty where SPL and KDE provide no 
uncertainty estimation.

\section{CDC flu forecasts analysis} \label{sec:qm_analysis}

In this section, the QGP and other models are 
used for QM of quantile forecasts targeting 
US flu hospitalizations. Beginning in 2013, the CDC began hosting a yearly 
forecast competition of the influenza outbreak in the US. This competition is 
known as FluSight. The flu epidemic typically begins in the fall and ends in 
late spring the following year, and the forecast competition lasts around 30 
weeks starting in October and ending in May. FluSight involves 
a few dozen academic 
and industry research teams who each independently develop forecasts every 
week for predicting certain flu targets for future weeks 
\cite[]{biggerstaff2016results}. During the 2022-23 and 2023-24 seasons, the 
forecast targets were the 1, 2, 3, and 4-week ahead hospitalizations as 
reported by Health and Human Services (HHS) for the 50 US states, Puerto Rico, 
the District of Columbia, and the nation as a whole. The data for weekly 
hospitalizations of flu patients may be found at \cite{healthdata2024covidts}. 
The official guidelines of the 2023-24 FluSight and all submitted quantile 
forecasts are publicly available at \cite{mathis2023flusight} 
\cite[]{mathis2024evaluation}.

\begin{figure}[hbt!]
    \centering
    \includegraphics[scale=.6]{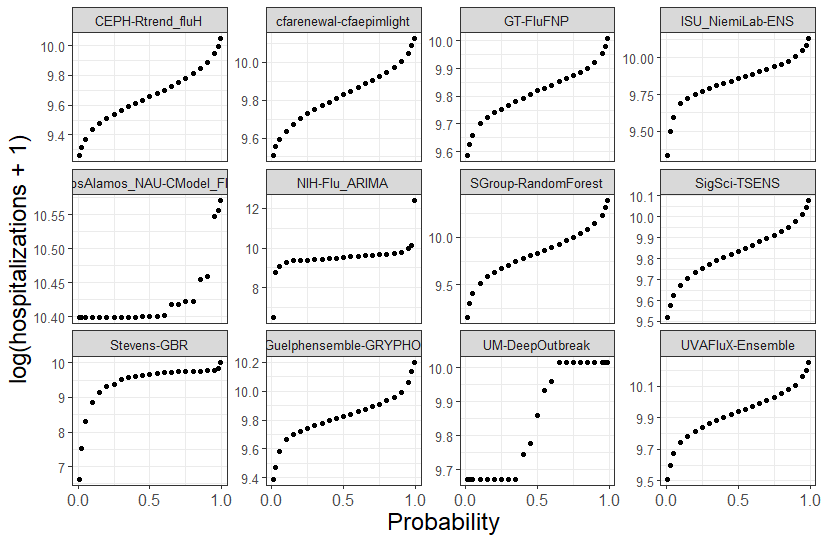}
    \caption{1 week ahead log flu hospitalization quantile forecasts from 12 
    teams who participated in the 2023-24 CDC flu forecast competition. 
    Forecasts are for the national level during the week of January 13, 2024.}
    \label{fig:quant_forcs_us}
\end{figure}

Participating teams were free to create forecasts however they pleased, but 
forecasts for each target were required to consist of 23 quantiles for 
probability levels 
$\boldsymbol{p} =(0.01, 0.025, 0.5, 0.1, ..., 0.95, 0.975, 0.99)$, 
which were given by FluSight. Figure \ref{fig:quant_forcs_us} shows examples 
of quantile forecasts of the same target from 12 participating teams. Each plot 
shows the log forecast for hospitalizations in the US for the week of 
January 13, 2024, and it is clear that there are major distributional 
differences between forecasts.
We denote a quantile forecast for flu hospitalizations as 
$\hat{Q}^{(H)}(\boldsymbol{p}) = (\hat{Q}^{(H)}(0.01), ..., \hat{Q}^{(H)}(0.99))$. 
The general quantile forecast representation format allowed for forecasts to be 
compared by the same metric, the weighted interval score (WIS), and to be 
easily combined into a multi-model ensemble forecast 
\cite[]{mathis2024evaluation}. However, under the quantile representation, 
the tools for scoring forecasts and for building ensemble models are limited 
as many of the existing tools for scoring forecasts and constructing ensembles 
require CDFs or PDFs \cite[]{wadsworth2023mixture,ranjan2010combining}. 
Because of these limitations, it may be desirable to approximate continuous 
distributions from the quantile forecasts to allow for more flexibility in 
scoring or ensemble building. In this section, we fit the QGP and other
models for QM to 
forcasts submitted to the FluSight during the 2023-24 
season, and an analysis is made to compare the fits.

Before fitting a QGP model to a quantile forecast we transformed the quantiles 
$\hat{Q}^{(H)}(\boldsymbol{p})$ to $\text{log}(\hat{Q}^{(H)}(\boldsymbol{p}) + 1)$ 
so that forecasts for all states were on a similar scale. The model fit was 
the same as the QGP fit in section \ref{sec:uk_matching}, including the same 
prior distribution assignments.
According to the official forecast competition rules, no forecast could include 
quantiles that were less than 0. As a result, there were forecasts with one or 
many quantiles equal to 0. When fitting the QGP model, the 0 
values were removed so that some forecasts had $K < 23$ quantiles. Between 39 
competing forecast teams, 53 locations, 29 forecast dates, and 4 horizons, 
there were 161,093 quantile forecasts on which QM was done.
For each 
forecast, the WIS was calculated for the quantile forecast and the continuous 
ranked probability score (CRPS) was calculated over the predictive distribution 
of the fit QM models.

The WIS and CRPS are both proper scoring rules, which is a class of scoring 
rule defined so as to keep a forecaster honest in their forecasts 
\cite[]{gneiting2007strictly, gneiting2014probabilistic}. 
The definition for the WIS is in (\ref{eq:wis}) and is the same as that in 
\cite{bracher2021evaluating}. The WIS consists of the sum of multiple intervals 
scores (IS), the definition of which is in (\ref{eq:is}). Here $\alpha$ is the 
nominal level of an interval with $l$ and $r$ being the respective lower and 
upper bounds of the interval. $R$ is the number of intervals in the quantile 
forecast, $\alpha_r$ is the nominal level for the $r^{th}$ interval, and $y^*$ 
is the observed value which one is attempting to forecast. 

\begin{equation}
\label{eq:wis}
        WIS_{0,R}(F, y^*) = \frac{1}{R + 1/2} \times 
        (w_0\times |y^* - median| + 
        \sum_{r=1}^R \{w_r \times IS_{\alpha_r}(F, y^*) \} )
\end{equation}

\begin{equation}
\label{eq:is}
        IS_{\alpha}(l,r;y^*) = (r-l) + 
        \frac{2}{\alpha}(l - y^*)\mathds{1}\{y^* < l\} + 
        \frac{2}{\alpha}(y^* - r) \mathds{1}\{y^* > r\}
\end{equation}

The CRPS is a widely used scoring rule for forecasting and is a function of the 
CDF of a continuous distribution thus making it unavailable for scoring 
quantile functions. The CRPS is defined in (\ref{eq:crps}) where $F$ is the 
CDF of a forecast and $y^*$ is the observed event one attempts to forecast. 
The definition is the same as in \cite{gneiting2014probabilistic}. 
\begin{equation}
    \label{eq:crps}
    \text{CRPS}(F, y^*) = \int_{-\infty}^{\infty} 
    (F(x) - \mathds{1} (y^* \leq x))^2 dx
\end{equation}
Fitting continuous distributions to the quantile forecasts would allow for 
forecast comparison using the CRPS. The CRPS assesses a forecast across an 
entire distribution, including the tails, which the WIS is unable to do, 
giving more reason why one may want to perform QM on a quantile forecast. 
We fit the QGP model in (\ref{eq:pit_qgp}) to the forecast competition 
forecasts from the 2023-24 season and compare the results of scoring the given 
quantile forecasts by the WIS and the CRPS calculated from the posterior 
predictive of the QGP model. Posterior predictive samples from the QM 
forecasts allow for approximate calculation of the CRPS. 
To calculate the WIS, we use the \texttt{evalcast} \texttt{R} package 
\cite[]{mcdonald2023evalcast}, and to calculate the CRPS we use the 
\texttt{scoringutils} package \cite[]{jordan2019scoringutils}.

\begin{figure}[hbt!]
    \centering
    \makebox[\textwidth][c]{%
        \adjustbox{trim=0 0 0 0,clip}{\includegraphics[scale=.5]{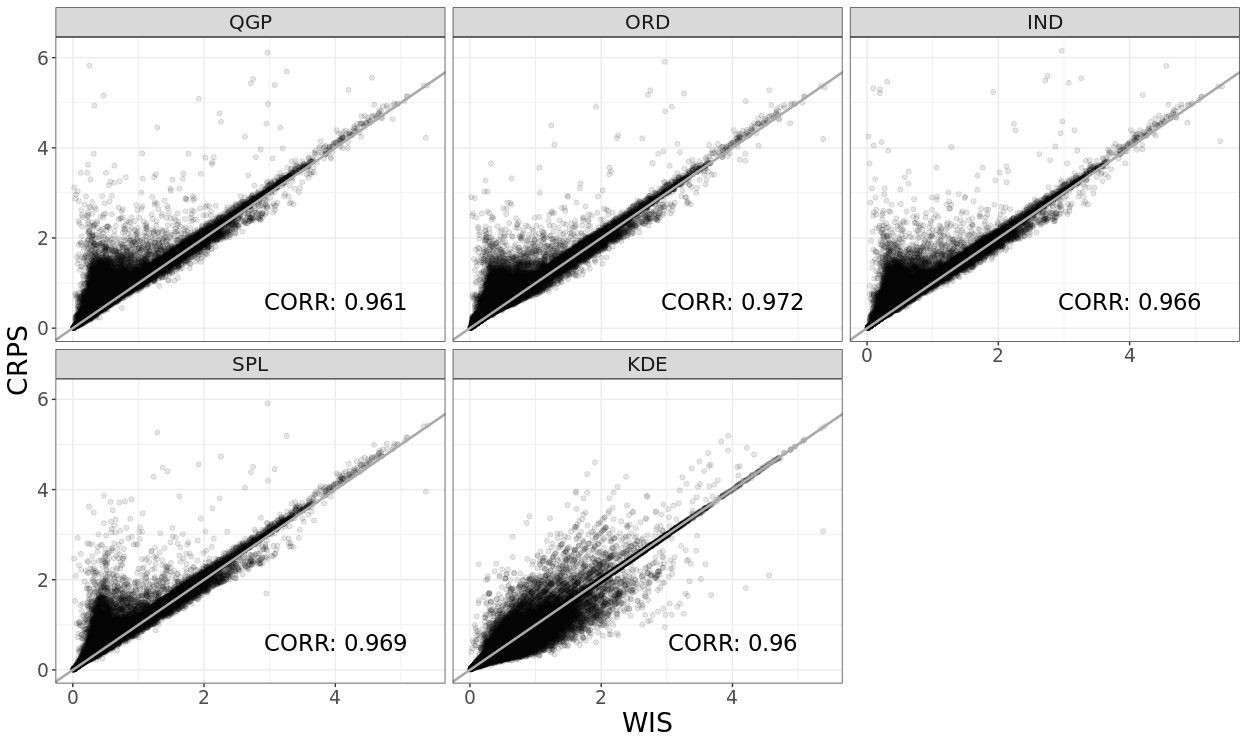}}%
    }
    \caption{Scatterplots of WIS and CRPS values for forecasts from 39 
    different teams for 2023-24 CDC flu forecast competition. Each plot
    is for a different QM method. Each point represents scores for a flu 
    forecast for one week during the season, one state, and one competing team. 
    Points are transparent to show where more scores tended to be. Overall 
    linear correlation is also given in the corner of each plot.}
    \label{fig:flu_methods_corr}
\end{figure}

Figure \ref{fig:flu_methods_corr} shows results of 
comparing the WIS from quantile forecasts and the CRPS from the 
fitted forecasts for QGP, ORD, IND, SPL, and KDE methods. 
The plots have in the $x$-axis the WIS and 
CRPS is in the $y$-axis. Each point represents
one forecast for any participating 
team, season week, and state. Unsurprisingly, the correlation between the
QM CRPS and WIS is very highly correlated for each method. For all but the KDE
method, there is a tendency for some CRPS values to go above the $x = y$ line
suggesting the CRPS scores are enlarged due to the addition of distribution
tails. The CRPS scores for the IND method also go below the $x = y$ line 
suggesting that the method's fit may be simply be poorer than the other methods.
The slightly lower correlation for the QGP compared to ORD and IND may also be
a result of adding more uncertainty in the fitting of tail quantiles.

\begin{figure}[hbt!]
    \centering
    \makebox[\textwidth][c]{%
        \adjustbox{trim=0 0 0 0,clip}{\includegraphics[scale=.45]{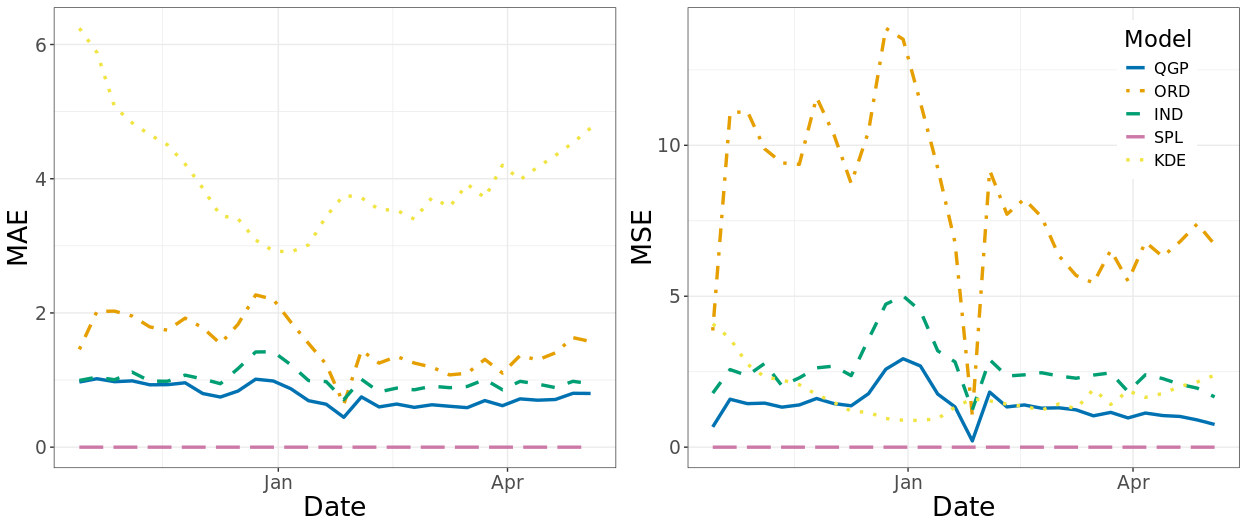}}%
    }
    \caption{Mean absolute error (MAE) and mean square error (MSE) over week
    for FluSight QM quantiles for each of QGP, ORD, IND, SPL, and KDE methods.}
    \label{fig:all_methods_dists}
\end{figure}

We do not know the true distributions from which each quantile forecast was
made, but we can assess closeness of fit by measuring the difference between
the QM quantiles and the original forecast quantiles given the FluSight
probability values.
Figure \ref{fig:all_methods_dists} shows the average over week mean absolute
error (MAE) and mean sqaure error (MSE) between the the newly estimated 
QM quantiles and the original forecast quantiles. The MAE and MSE values for the
SPL method are 0 for every week because the estimated quantile functions are 
interpolations of the original forecasts. Besides the SPL, however, the QGP
appears to estimate quantiles that are the closest to the original forecast
quantiles. The average MAE for the QGP is smaller than for the remaining methods
for each week and the MSE is smaller for most weeks showing a robustness which
the other methods -- namely ORD and KDE -- may lack.

\section{Conclusion} \label{sec:qm_conclusion}

In this manuscript
we considered the situation where quantiles with known probability 
levels are given as data or the quantiles represent a distribution rather than 
raw data or a fully defined distribution. We reviewed basic properties of QFs, 
sample quantiles, and resulting CLTs which motivate a new model, the QGP 
model, for matching a set of quantiles to a continuous distribution. The QGP 
model is assessed for parameter estimation and inference, distribution 
approximation, and it is compared with other QM methods found in the 
literature. The QGP provides explicit and easily interpretable uncertainty 
quantification of QM distributions, and generally outperforms the methods to 
which it is compared, sometimes with the exception of the ORD.
An added advantage, however, is that the QGP 
provides a way to perform QM for distributions defined by CDFs as well as 
quantile distributions where the CDF is difficult to evaluate.
An application of QM on quantile forecasts from the 2023-24 
FluSight project demonstrates that QM methods allow for the use of the CRPS
for scoring forecasts which is able to account for distribution tail 
uncertainty. 
QGP and other QM forecasts may be scored using a variety of scoring rules 
unavailable to quantile forecasts, and independent forecasts may be combined 
using methods made only for combining distributions by CDF or PDF functions. 
QGP has already been applied in chapter 4 of \cite{wadsworth2024quantile} who 
used the QGP predictive distributions fit in section \ref{sec:qm_analysis} to 
construct ensemble forecasts.
In the analysis, it is shown that the QGP may offer a more robust
fit to quantile distributions than other QM methods.

Approximating continuous distributions given only a set of estimated quantiles 
is not a new idea, but QGP provides a method of doing so while also accurately 
accounting for asymptotic uncertainty in estimated or predicted quantiles. The 
CLTs in this manuscript
are limited to the case where the quantiles are estimated 
given a distribution sample, however similar results have been shown for the 
quantile regression case where quantiles are estimated in the presence of 
covariates \cite[]{kocherginsky2005practical, koenker1978regression}. These 
results could lead to additional QM modeling using the QGP model where 
covariates are given, and this may be more akin to the QM done by 
\cite{sgouropoulos2015matching}. 
Additional research for the QGP model may be in how the required distribution 
function is selected. Where a true model is unknown, we elected to model a 
distribution as a normal mixture distribution. Of course this has its limits, 
and using other non parametric functions, similar to 
\cite{gasthaus2019probabilistic}, may provide needed flexibility. As long as 
the selected function is continuous and once differentiable, the theory herein 
applies. 
The applications here suggests QM being useful for forecast hubs.  
\cite{gerding2023evaluating} found QM a necessity in order to evaluate 
forecasts by a new scoring rule with COVID-19 specific applications.

\section*{Acknowledgments}
This work is partially supported by the National Science Foundation under 
Grant No. 2152117. Any opinions, findings, and conclusions or recommendations 
expressed in this material are those of the author(s) and do not necessarily 
reflect the views of the National Science Foundation.

\section*{Data Availability}
All data used in this manuscript may be found 
at \href{https://github.com/cdcepi/FluSight-forecast-hub}{https://github.com/cdcepi/FluSight-forecast-hub}

All code used to obtain results in this manuscript, including code for 
accessing data, may be found at 
\href{https://github.com/wadspen/quantile_fitting}{https://github.com/wadspen/quantile\_fitting}

\appendix


\bibliographystyle{unsrt}
\bibliography{master_bib}



%
%
%
\newpage
%
%
%
%
%
%

\title{Supplementary Material} 

\Huge
\begin{center}
  Supplementary Material
\end{center}
\normalsize

\section*{Additional simulation study results to section 4.1}
\begin{figure}[hbt!]
\centering
  \includegraphics[width = 1\linewidth]{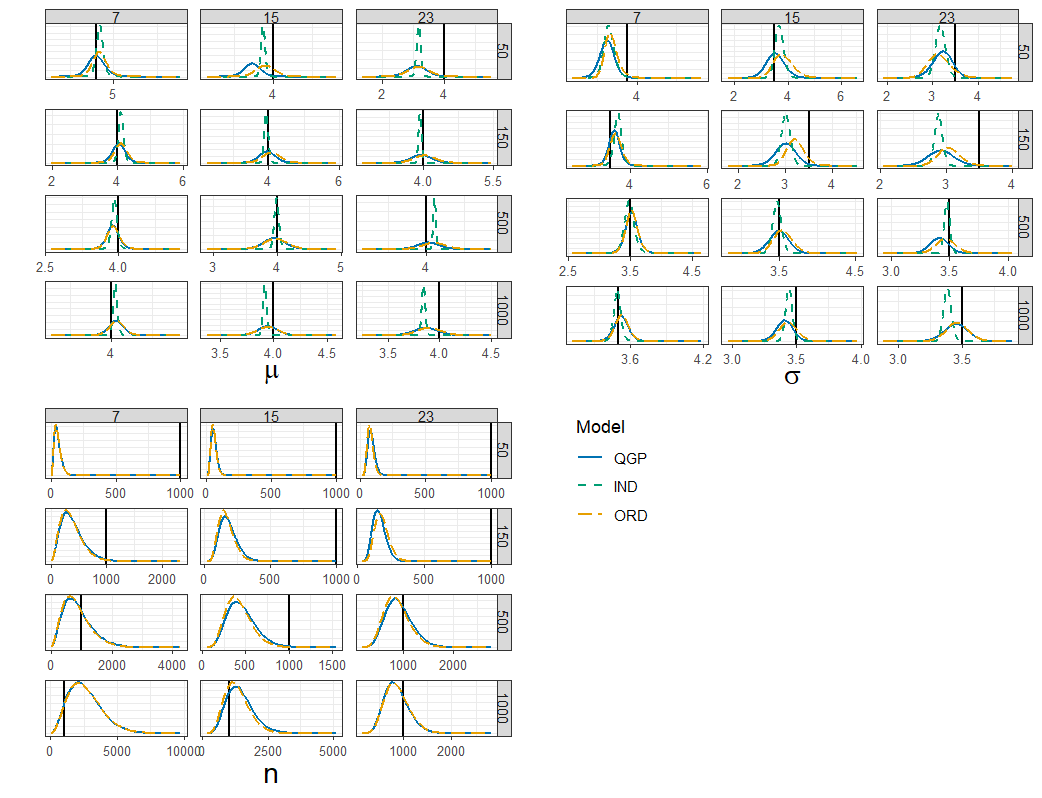}
\caption{Density plots of posterior distribution samples for normal parameters 
by QM for QGP, ORD, and
IND models. QM was done on estimated quantiles from a normal distribution with 
mean 4 and standard
deviation 3.5. The posterior densities are for $\mu$ (top left), 
$\sigma$ (top right), and 
sample size n (bottom). Plots are
faceted by the sample size $n \in \{50, 150, 500, 1,000\}$ 
($y$-axis) and number of 
quantiles $K \in \{7, 15, 23\}$ ($x$-axis).
Vertical lines (black) show the value of the true parameter}
\label{fig:norm_dens}
\end{figure}

Figure
\ref{fig:norm_dens} shows examples of marginal
posterior densities for QM of quantiles for a normal distribution from the
simulation study in section 4.1 in
the main manuscript. The parameter uncertainty of the QGP and ORD models are
unsurprisingly similar in
most cases, and where they differ it is hard to say that one is estimating the
true parameter better than the
other. The IND model on the other hand has tighter densities, and often the
bulk of the posterior is far from
the true parameter.

\begin{figure}[hbt!]
\centering
  \includegraphics[width = 1.1\linewidth]{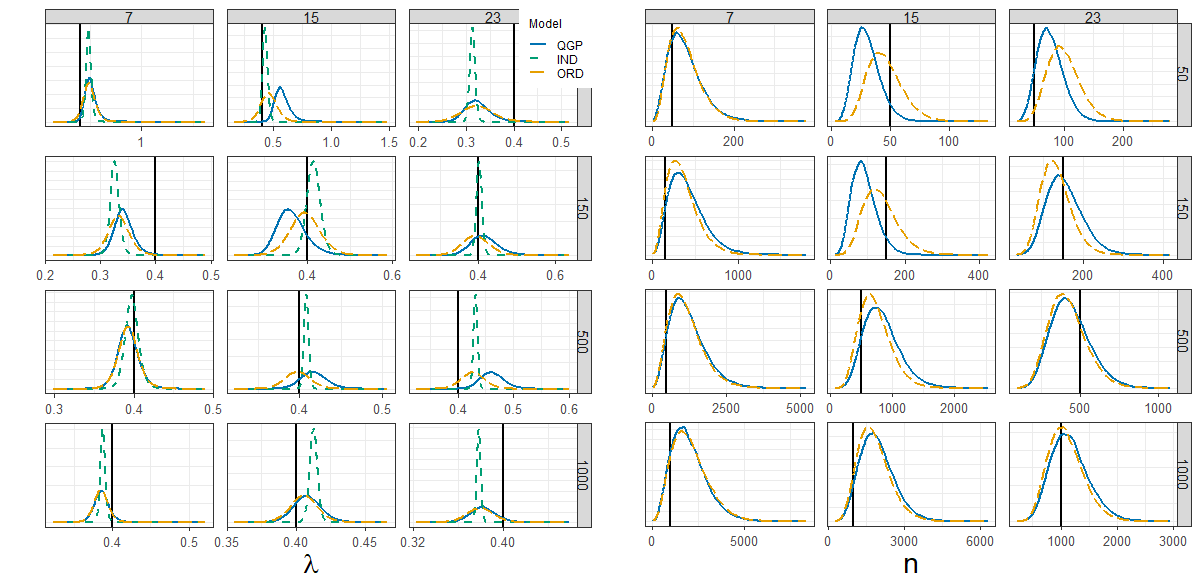}
\caption{Density plots of posterior distribution samples for the exponential 
parameters by QM for QGP,
ORD, and IND models. QM was done on estimated quantiles from a exponential 
distribution with parameter
$\lambda = 4$. The posterior densities are for $\lambda$
(left) and sample size $n$ (right). 
Plots are faceted by true sample
size $n \in \{50, 150, 500, 1,000\}$ ($y$-axis) and number of quantiles 
$K \in \{7, 15, 23\}$ 
($x$-axis). Vertical lines (black)
show the value of the true parameter.}
\label{fig:exp_dens}
\end{figure}

\begin{figure}[hbt!]
  \includegraphics[width=\linewidth]{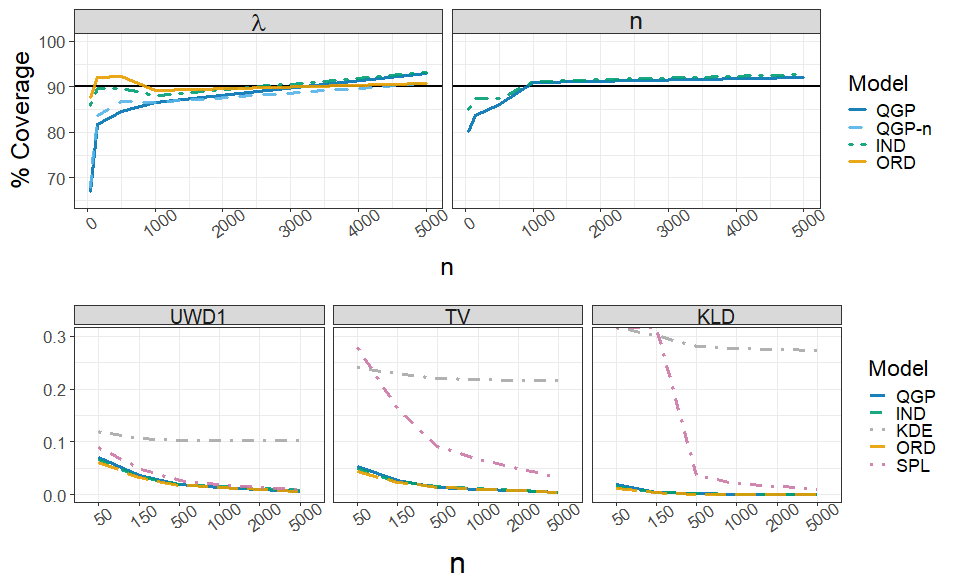}

\caption{ Posterior coverage (top) calculated as the percentage of times the 
true parameter fell within the
modeled 90\% credible interval over the 500 replications. Coverage is faceted 
by the exponential parameter $\lambda$
and $n$ with $K = 23$, and by increasing sample size ($x$-axis). 
The five models 
QGP, ORD, QGPN, ORDN,
and IND are colored as shown the legend. The horizontal line (black) is at 
the nominal 90\% level. Only QGP
and ORD appear for the parameter $n$ as they are the only two which estimate 
an unknown $n$. (bottom)
Distance between the true distribution and the estimated QM predictive 
distribution averaged over the 500
replications. Distances include UWD1, TV, and KLD for $K = 23$, 
and by increasing sample size (x-axis)}
\label{fig:exp_cov_dists}
\end{figure}

\newpage
Figures \ref{fig:exp_dens} and \ref{fig:exp_cov_dists} are from a simulation 
study similar to that of section 4.1 
except that QM was performed on
quantiles estimated from draws from an exponential distribution with 
parameter $\lambda = 4$ rather than a normal distribution. 
The results in the study for the exponential distribution were similar 
to those for the normal
distribution.

\section*{Additional simulation study results to section 4.2}

\begin{figure}[hbt!]
\centering
  \centering
  \makebox[\textwidth][c]{%
        \adjustbox{trim=0 0 0 0,clip}{\includegraphics[width=1.3\textwidth]{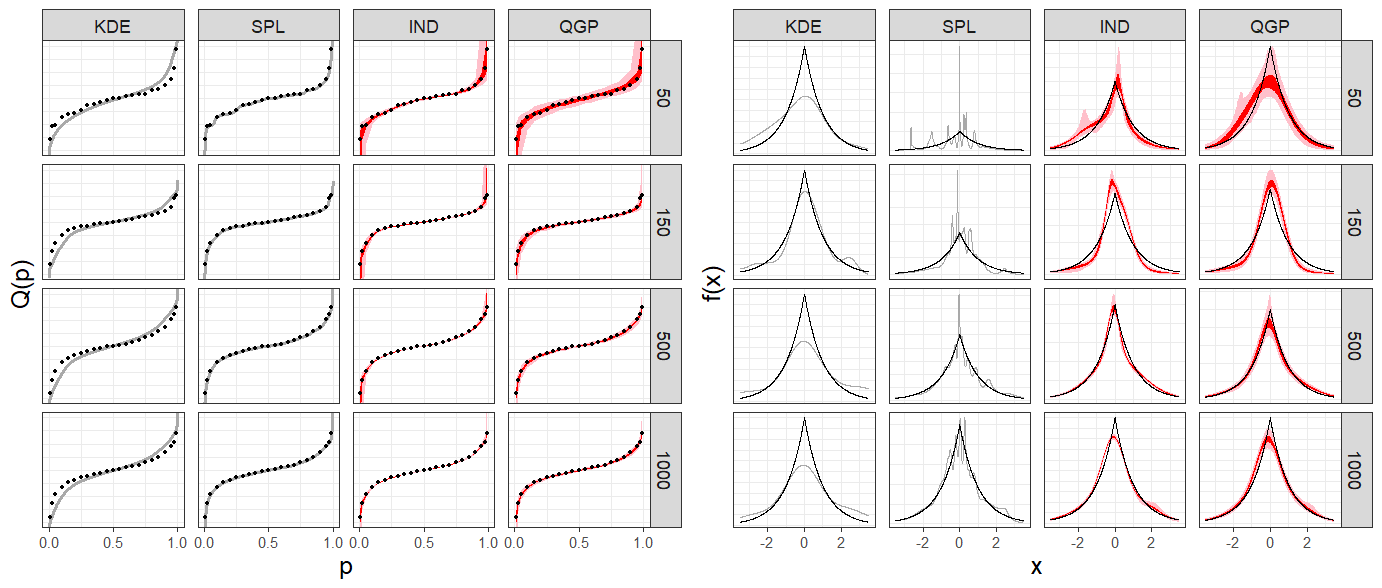}}%
    }

\caption{QM fits of $K=23$ quantiles by KDE, SPL, IND, and QGP for $n \in \{50, 150, 500, 1{,}000\}$. The quantiles were sampled from the Laplace distribution $La(0,1)$. The quantile fits (left) show the true quantiles (black) with either the QM fit line (grey) or the credible intervals of 50\% (red) and 95\% (pink). 
The estimated PDF plots (right) show the true PDF (black) with either a the QM estimated PDF (grey) or the credible intervals of 50\% (red) and 95\% (pink).}
\label{fig:lp_fits}
\end{figure}

\begin{figure}[hbt!]
\centering
  \centering
  \makebox[\textwidth][c]{%
        \adjustbox{trim=0 0 0 0,clip}{\includegraphics[width=1.3\textwidth]{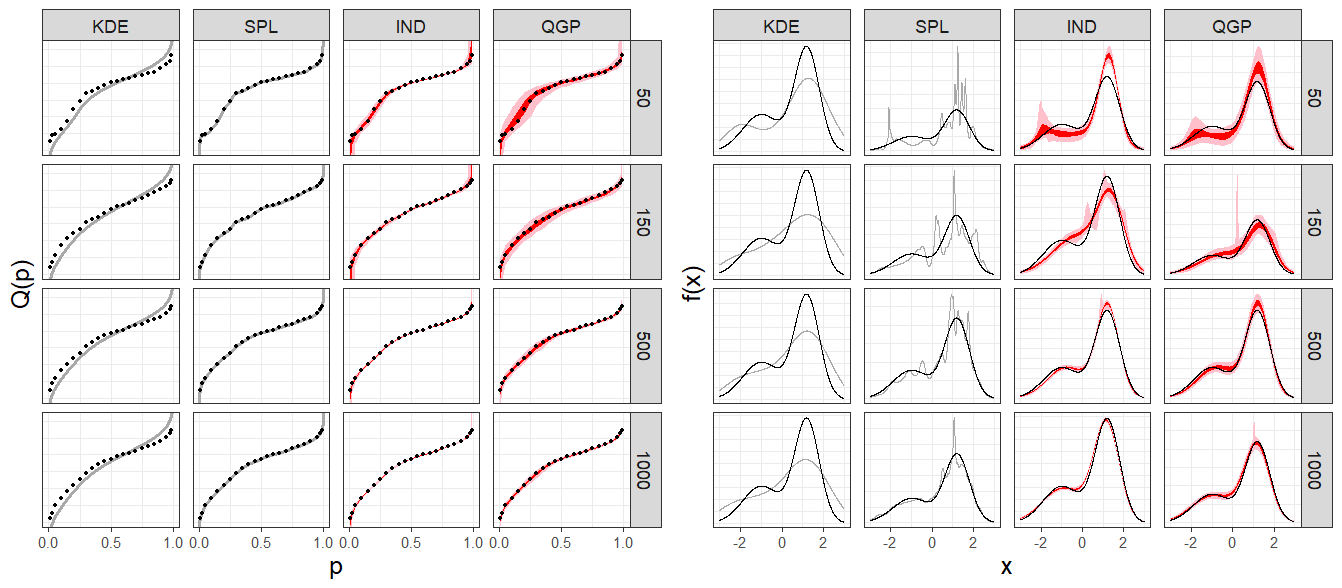}}%
    }
\caption{QM fits of $K=23$ quantiles by KDE, SPL, IND, and QGP for $n \in \{50, 150, 500, 1{,}000\}$. The quantiles were sampled from the two component normal mixture distribution $w N(-1, 0.9) + (1-w)N(1.2, .6)$ where $w = 0.35$. The quantile fits (left) show the true quantiles (black) with either the QM fit line (grey) or the credible intervals of 50\% (red) and 95\% (pink). 
The estimated PDF plots (right) show the true PDF (black) with either a the QM estimated PDF (grey) or the credible intervals of 50\% (red) and 95\% (pink). }
\label{fig:gmix_fits}
\end{figure}

Figures \ref{fig:lp_fits} and \ref{fig:gmix_fits} show examples of QM fits
to Laplace and normal mixture distributions as described in section 4.2.

\newpage

\section*{QGP normal mixture components analysis}

\begin{figure}[hbt!]
\centering
  \includegraphics[width=.9\textwidth]{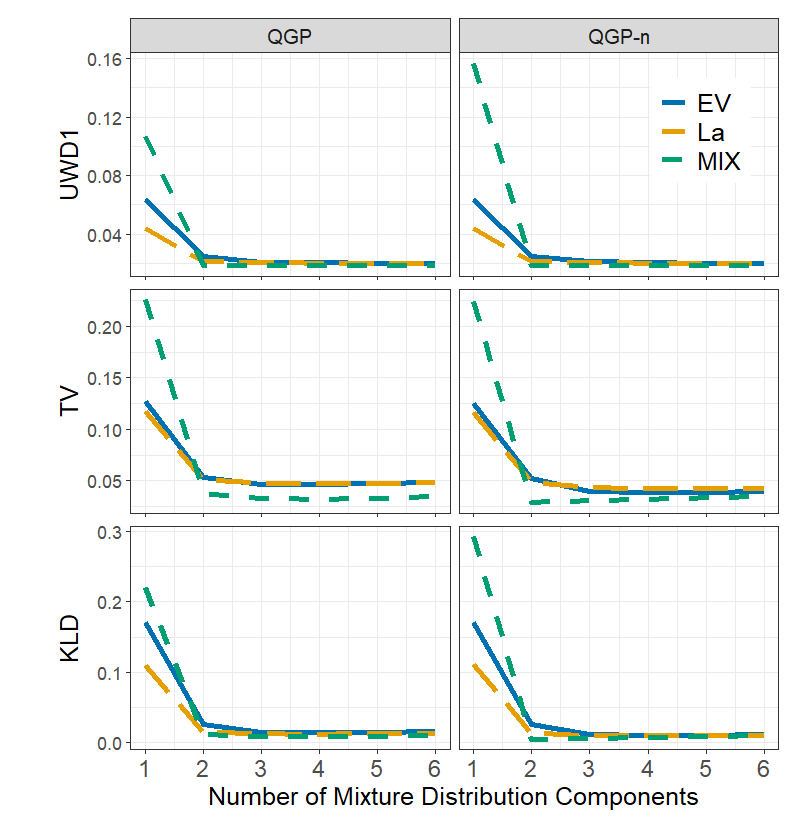}
\caption{UWD1, TV, and KLD distances averaged over 500 simulation replicates
for QGP fits of quantile simulated from extreme value, Laplace, and
mixture distributions. The QGP with normal 
mixture distributions of 1 to 6 components were fit for both the cases where 
$n$ is not known (left) and $n$ is known.}
\label{fig:mix_comps}
\end{figure}

This section shows an analysis which was made to determine the number of 
normal components to use in the distribution assigned for use in the QGP.
Figure \ref{fig:mix_comps} shows the UWD1, TV, and KLD distances between 
a QGP fit distributions and simulated quantiles averaged over 500 simulated
replicates. Table \ref{tab:mix_comps} shows the average values. The results 
show that while the larger the number of components, the closer is the fit 
distribution in terms of the given distances. However, the imrpovement after 
three or four components is not much making four components attractive 
for relatively good fit and for model simplicity.

\newpage

\begin{table}[hbt!]
\footnotesize
\centering
\caption{UWD1, TV, and KLD distances averaged over 500 simulation replicates
for QGP fits of quantile simulated from extreme value, Laplace, and
mixture distributions. The QGP with normal 
mixture distributions of 1 to 6 components were fit for both the cases where 
$n$ is not known (left) and $n$ is known.}
\begin{adjustbox}{center}
\begin{tabular}{clcccccc|cccccc}
& & \multicolumn{6}{c}{QGP} & \multicolumn{6}{c}{QGP-n} \\
\cmidrule(lr){3-8} \cmidrule(lr){9-14}
& Components & 1 & 2 & 3 & 4 & 5 & 6 & 1 & 2 & 3 & 4 & 5 & 6 \\
\midrule
\multirow{3}{*}{UWD1} &
EV & 0.063 & 0.025 & 0.021 & 0.020 & 0.020 & 0.020 & 
      0.064 & 0.025 & 0.021 & 0.021 & 0.020 & 0.020\\
 & La & 0.044 & 0.022 & 0.020 & 0.020 & 0.020 & 0.020 & 
      0.044 & 0.021 & 0.020 & 0.020 & 0.020 & 0.020 \\
 & MIX & 0.107 & 0.018 & 0.019 & 0.019 & 0.019 & 0.019 & 
      0.157 & 0.018 & 0.019 & 0.018 & 0.018 & 0.018 \\
      \hline
\multirow{3}{*}{TV} &
EV & 0.127 & 0.053 & 0.047 & 0.047 & 0.048 & 0.049 &
      0.125 & 0.052 & 0.039 & 0.039 & 0.039 & 0.039\\
 & La & 0.117 & 0.051 & 0.047 & 0.047 & 0.048 & 0.048 &
      0.117 & 0.048 & 0.043 & 0.043 & 0.043 & 0.042 \\
 & MIX & 0.226 & 0.038 & 0.033 & 0.032 & 0.033 & 0.034 &
      0.225 & 0.029 & 0.031 & 0.032 & 0.034 & 0.035 \\
      \hline
\multirow{3}{*}{KLD} &
EV & 0.171 & 0.025 & 0.014 & 0.013 & 0.014 & 0.015 & 
      0.170 & 0.025 & 0.011 & 0.010 & 0.010 & 0.010\\
& La & 0.110 & 0.014 & 0.012 & 0.012 & 0.012 & 0.012 & 
      0.111 & 0.013 & 0.010 & 0.010 & 0.010 & 0.010 \\
& MIX & 0.221 & 0.010 & 0.008 & 0.008 & 0.009 & 0.010 & 
      0.293 & 0.004 & 0.006 & 0.007 & 0.009 & 0.010
\end{tabular}
\end{adjustbox}
\label{tab:mix_comps}
\end{table}





\end{document}